\documentclass[journal]{IEEEtran}
\usepackage{graphicx}
\usepackage{amsmath}
\usepackage{amssymb}
\usepackage{booktabs}
\usepackage{algorithm}
\usepackage{algorithmic}
\usepackage{subfig}
\usepackage{algorithm}
\usepackage{amsthm}
\usepackage{placeins}

\theoremstyle{definition}
\newtheorem{definition}{Definition}[section]

\begin{document}

\title{Statistical Inference Attack Against PHY-layer Key Extraction and Countermeasures}

\author{Rui~Zhu,
        Tao~Shu,
        and~Huirong~Fu
\thanks{Rui Zhu and Huirong Fu are with the Department of Computer Science and Engineering, Oakland University, Rochester, MI, 48309 USA (email: {rzhu, fu}@oakland.edu)}
\thanks{Tao Shu is with the Department
of Computer Science and Software Engineering, Auburn University, Auburn,
AL, 36849 USA (e-mail: tshu@auburn.edu)}
\thanks{}}


\maketitle

\begin{abstract}
With the fast growth of high-performance computing, the security of traditional cryptographic secret key establishment mechanisms are seriously challenged by computing-intensive attacks. As an alternative, considerable efforts have been made to develop physical (PHY) layer security measures in recent years, such as link-signature-based (LSB) secret key extraction techniques. Those mechanisms have been believed secure, based on the fundamental assumption that wireless signals received at two locations are uncorrelated when they were separated by more than half a wavelength apart. However, this assumption does not hold in some circumstances under latest observations, rendering LSB key extraction mechanisms vulnerable to attacks. The formal theoretical analysis on channel correlations in both real indoor and outdoor environments are provided in this paper. Moreover, this paper studies empirical statistical inference attacks (SIA) against LSB key extraction, whereby an adversary infers the signature of a target link. Consequently, the secret key extracted from that signature has been recovered by observing the surrounding links. Prior work assumes theoretical link-correlation models for the inference, in contrast, our study does not make any assumption on link correlation. Instead, we take machine learning (ML) methods for link inference based on empirically measured link signatures. ML algorithms have been developed to launch SIAs under various realistic scenarios. Our experimental results have shown that the proposed inference algorithms are still quite effective even without making assumptions on link correlation. In addition, our inference algorithms can reduce the key search space by many orders of magnitudes compared to brute force search.
We further propose a countermeasure against the statistical inference attacks, FBCH (forward-backward cooperative key extraction protocol with helpers). In the FBCH, helpers (other trusted wireless nodes) are introduced to provide more randomness in the key extraction. Our experiment results verify the effectiveness of the proposed protocol.

\end{abstract}


\section{Introduction}
Secret key provides confidentiality and integrity in communication.
The establishment of secure secret keys ahead of transmissions is
one of the key issues in the field of information security. The
Diffie-Hellman key agreement protocol (1976) was the first practical
method for establishing a shared secret over an unsecured
communication channel. The security of Diffie-Hellman protocol is
based on the discrete logarithm problem, whose solution is assumed
to be hard to compute. However, with the fast growth of
high-performance computing, the above assumption is seriously
challenged, rendering Diffie-Hellman possibly vulnerable to various
computation-intensive attacks.

Realizing the potential vulnerability of Diffie-Hellman, developing
new key extraction mechanisms whose security do not rely on the
computation hardness assumption receives many attentions. One
solution is through PHY-layer security, which extracts
symmetric secret keys from the PHY-layer channel response of
the wireless link (i.e., the link signature) between the transmitter
and the receiver, e.g.,
see~\cite{1203156,Patwari:2007:RLD:1287853.1287867,5462231,4291564}. The
channel response or link signature is considered to be a good pick
for secure key establishment because it is both reciprocal and
uncorrelated. It is reciprocal because it is usually assumed that
when the transmitter and receiver, Alice and Bob, make measurements
on the channel state of the link between them, their measurements
are symmetric (identical). On the other hand, the channel is said to
be uncorrelated because it is usually assumed that the state of any
other link separated by at least half of a wavelength apart from
Alice and Bob should be independent from that of the link between
Alice and Bob~\cite{1212671}. Based on these assumptions, it is
commonly believed that common keys can be extracted by Alice and Bob
based on their symmetric observation of the channel between them,
while this channel is unobservable by a third party separated far
enough (half a wavelength) from Alice and Bob, making the extracted
keys secure and secrete.

While PHY-layer secret key extraction has been used in many
applications such as encryption and authentication, recent studies
have revealed that the uncorrelation assumption between separated
links may not always be valid~\cite{6566763,7321830}, especially in
many indoor environments where radio propagation becomes complicated
due to signal reflection and multipath. This opens the door for the
statistical inference attack (SIA) against the link-signature based
(LSB) key extraction, because the correlation between links may be
exploited by an adversary to probabilistically infer the signature of
a target link based on observations over surrounding links. In light
of such a vulnerability, SIA against LSB key extraction has been
analytically studied in prior work, by assuming a correlation model
between neighboring links,​ e.g.,
see~\cite{6566763,7321830,Edman:2011:PIA:1972551.1972559}. However,
it remains to be seen that, in a realistic wireless environment,
without making assumptions on the link correlation model, how and to
what extent SIA may undermine the security strength of LSB key
extraction.

In this paper, we explore the answer to the above questions by
taking a ML approach. We first discuss the correlation between two wireless channels in both indoor and outdoor environments. We build two models, indoor and outdoor communication models. According to these two models, the formal theoretical analysis on the channel correlation is provided. From the observation of theoretical analysis, we can find that there are still relative strong correlations even two links are separated further than half wavelength. Thus, it is possible for the adversary to launch correlation attacks by using the correlation information between the legitimate links and surrounding links. Our inference attack against existing key extraction scheme study roots from empirically
measured channel data and does not rely on any assumption on the
link correlation model. In particular, our study is based on the
CRAWDAD dataset~\cite{Patwari:2007:RLD:1287853.1287867}, which
contains over 9300 measured channel traces for 1892 links in a
44-node indoor office-type wireless network. Several possible SIA
scenarios are considered. For each scenario, measured link
signatures in CRAWDAD are divided into two datasets: training data
and test data. ML-based channel inference algorithms are
developed. We start our study from establishing artificial neural networks (ANN) models for inference.  After being trained based on the training data, these
algorithms are instructed to infer the link signatures in the test
dataset. Then, we utilize different ML algorithms, such as ensemble methods, support vector machine (SVM), and multivariate linear regression to launch SIA, and compare the inference performances of different ML algorithms. Our experiment results show that all these ML algorithms have approximative inference performance, and
thus can effectively reduce the key search space by many orders of
magnitudes compared to a brutal-force search mechanism.

In light of the above vulnerability, we propose a novel multi-link
Forward-backward Cooperative Key Extraction Protocol with Helpers
(FBCH) in this paper as a countermeasure to the aforementioned SIA
attacks, aiming to make the LSB key extraction more secure. In
particular, by introducing a set of helpers (these are legitimate
nodes assisting the key extraction process), FBCH allows two
communicating terminals to extract symmetric secret keys based on
the combined channel impulse responses (CIRs) of several randomly selected links. This is in
sharp contrast to the conventional method where the key extraction
is only dependent on the particular link between the transmitter and
the receiver. Consequently, the resulting key extraction becomes
less dependent on a particular fixed channel, making the
aforementioned SIA attacks, which mainly target the channel between
the two communicating terminals, less effective.

As a summary, the main contributions of this work are three-fold:
(1) We theoretically verify the existence of correlation between
neighboring links in realistic environments, (2) we suggest
empirical methods to exploit the correlation to launch SIA against
LSB key extraction, and (3) we further propose a countermeasure to
weaken the effects of SIA attacks and make LSB key extraction more
secure. To the best of our knowledge, this is the first systematic
and empirical study of LSB key extraction from the SIA perspective
in the literature.

Part of this work has been presented previously as a conference
paper in ~\cite{Zhu2017EmpiricalSI}. In contrast to our prior
conference paper, the journal version provides a formal theoretical
verification for the correlation between neighboring links, and also
proposes the FBCH countermeasure to the SIA attacks as well.

The rest of this paper is organized as follows. We review related
work in Section~\ref{sec:related}. Section~\ref{sec:mod} presents
the background of LSB key extraction schemes and
defines the system model. Section IV analyzes the correlation
between two wireless channels in both indoor and outdoor
environments. We describe the proposed neural network based SIA
attacks in Section~\ref{sec:attack}. Section~\ref{sec:results}
evaluates the performance and effectiveness of the proposed attacks.
Section~\ref{sec:countermeasures} presents the FBCH countermeasures
to the inference attacks and we conclude our work in
Section~\ref{sec:conclusion}.

\section{Related Work}\label{sec:related}
The idea of exploiting wireless channel characteristics for
generating secret keys has received considerable attention in recent
years. A variety of LSB PHY-layer key
extraction schemes has been proposed, e.g,,
\cite{Mathur:2008:RES:1409944.1409960,6226870,7172523, 7414040,
7120014, 7417477, 5756230}. While the security of LSB key extraction
relies heavily on the uncorrelation assumption of channels, the
validity of this assumption has not been evaluated/verified in these
works.

Theoretical analysis on the correlation among links is conducted in
\cite{6566763,7321830}. In particular, these works derive
theoretical link correlation by taking into account the
spatial/geometric relations among the transmitter, receiver, and
signal reflectors. One ring model and the Daggan-Rapport model are
employed to derive the link correlation models under various
scenarios. Their main finding is that the
uncorrelation-beyond-half-wavelength assumption is not always valid,
and therefore it is necessary to use larger guard zones around the
transmitter and the receiver for secure LSB key extraction. Unlike
\cite{6566763,7321830}, our work derives theoretical link
correlation in general indoor and outdoor environments with the random distributed scatterers.

The research
topic of attacks against PHY-layer-based key extraction systems currently receives limited research input. Some researchers have reported that the current key
extraction schemes are vulnerable to passive eavesdropping
\cite{Edman:2011:PIA:1972551.1972559,Steinmetzer:2015:LPL:2766498.2766514} as well as active attacks \cite{6140616, Law:2009:ELJ:1464420.1464426, 6716049,6415588}.

Experimental study on inference attack against PHY-layer key
extraction is considered in
\cite{Edman:2011:PIA:1972551.1972559}, where the key extraction is
based on the value of a received signal strength indicator (RSSI).
Inferences in \cite{Edman:2011:PIA:1972551.1972559} are mainly based
on simple averaging methods and the RSSI observations used for the
inference are made close to the target link (ranging from 6 cm to at
most 90 cm away from the target receiver). In reality, making
inference attack based on overhearing the target channel at such a
close distance may not be practical. Our work considers a
significantly different inference model. In particular, instead of
inferring RSSI, we infer the channel response, which allows faster
key extraction. Furthermore, rather than being a simple averaging of
samples, our inference is based on neural network algorithms, which
enables a much better inference outcome through training. As a
result, our methods support inference based on observations made
much further away from the target link, ranging from meters to over
ten meters, which is of more interest in practice.  \cite{Steinmetzer:2015:LPL:2766498.2766514} also focuses on the passive attacks. It introduces a new analysis scheme that distinguishes
between jammed and unjammed transmissions based
on the diversity of jammed signals.

Attacks other than SIA have also been considered in the literature.
For example, \cite{6195669} studies the mimicry attack, where an
adversary replays or forwards legitimate responses from the
transmitter to the receiver. Countermeasures proposed include time
synching responses for verification and randomizing the training
sequence.

In \cite{6716049}, Zhou et al.  consider the active attacker scenario. They assume that Eve is active and can send attack signals to minimize the key extraction rate of current key extraction scheme. Eve's optimal attack strategy is characterized in \cite{6716049}. And they propose a scheme for the key extraction in the two-way relay channel. Instead of estimating CIRs by two terminals, Alice and Bob, the relay will first establish two pair-wise keys with Alice and Bob. Then the relay broadcasts the XOR of these two
pair-wise keys to both Alice and Bob. Alice and Bob can then
decode both keys and pick the one with a smaller size as the
final key. This scheme can effectively prevent against the active attacks. However, it requires the additional help from relays.

\cite{7346835} presents a new form of highly threatening active attack, named signal injection attack. The attacker can inject the similar signals to both two terminals to manipulate the channel measurements and compromise a portion of the key. PHY-UIR as a countermeasure to the signal injection attack is proposed in this work. In PHY-UIR, both two terminals introduce randomness into the channel probing frames. Thus, the random series that are used to extract secret keys are the combination of randomness in the fading channel and the ones introduced by users. Then the composed series are uncorrelated to the injected signals. As a result, the attacker is not able to compromise the composed secret keys. \cite{6415588} presents a formal active adversary model which takes into account
an adversary’s knowledge/control of the wireless channel.

We propose SIA attacks without
making specific assumption on link correlation. Rather than being a
theoretical study, we consider our work empirical/experimental, as
we use ML algorithms to make inference based on
empirically measured link signatures in a realistic environment.
Moreover, the goals of our work include not only verifying the
existence of correlation between nearby links, but also exploiting
such correlation for secret key inference through a set of
practically usable methods.

On a different track other than security, channel
inference/estimation has been extensively studied for efficient
radio resource management in wireless networks, e.g., for MIMO
systems~\cite{1203156, 6144766}. However, such inference/estimation
is made only for the channel between the target transmitter and the
receiver, rather than for channels beside the target link.

\section{Model Description}\label{sec:mod}
\subsection{Multi-path Effect and Link Signature}

In wireless communications, radio signals generally reach the receiving antenna by two or more paths due to reflection, diffraction, and scattering, which is called multipath propagation. Since different paths have different distances between transmitter and receiver, a receiver usually receives multiple copies of the transmitted signal at different time. Different copies have different attenuations due to the different path losses. The received signal is the sum of these delayed signal copies.

A radio channel consists of multiple paths from a transmitter to a receiver, and each path of the channel has a response (e.g., distortion and attenuation) to the multipath component traveling on it, which is called a component response. The superposition of all component responses is the \emph{channel impulse response} (CIR). Since the multi-path effects between different pairs of nodes, as well as channel impulse responses, are usually different, a channel impulse response between two nodes is also called a \emph{link signature}. 

\subsection{Key Extraction from Link Signature}
\begin{figure}[htbp]
	\centering
	\includegraphics[scale=0.3]{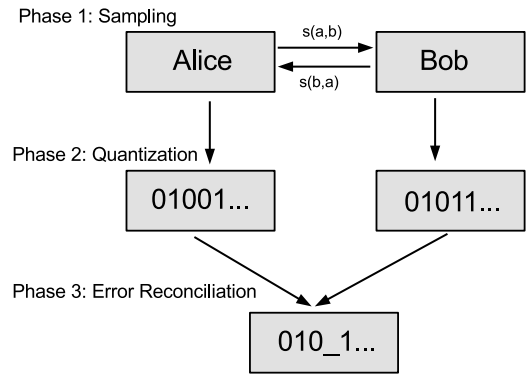}
	\caption{Framework to extract key bits from link signature.}\label{framework}
\end{figure}

Once channel impulse responses have been estimated, the process of
key extraction is rather straightforward.
First, channel impulse responses should be quantized for secret key
extraction since they are continuous random variables. There are
several kinds of mechanisms to quantize link signatures. In this
paper, we adopt the relative simple but widely-used uniform
quantization to quantize CIR measurements [3]. First, we normalize each
CIR with its maximum element value to obtain vectors of discrete decimals. Next, the resulting discrete decimals are multiplied with 32 and then are rounded to the nearest integers. In this way, we obtain vectors of integers in the range of [0,31], which are the quantization results of continuous channel impulse responses in integer representation. Then, the vectors of integer are converted to their binary presentation. Lastly, N-bit binary string is cut out from the whole string as the initial N-bit secret key. Figure~\ref{framework} shows the framework to extract key bits.

However, the straightforward quantization mechanism usually is not sufficient. Due to the variation of real environment and hardware differences between two measurement devices, it does not guarantee that the pairwise measurements from two communication ends (Alice and Bob) are identical. In this case, the sequences of bit keys extracted will not be identical. Therefore, we should employ an error reconciliation mechanism to solve this problem. For example, we can apply challenge-response verification protocol. Let $K_{A}$, $K_{B}$ are the bit keys extracted by Alice and Bob, respectively, and $\phi$ is a random number that Alice picked. To launch the verification protocol, Alice encrypts $\phi$ by her secret key $K_{A}$, and sends Bob
$E_{K_{A}}(\phi)$ and Bob responds with $E_{K_{B}}(\phi + 1)$. If
Alice gets $\phi + 1$ after she decrypts Bob's message, she can
conclude that Bob obtains the correct key. Bob can do likewise. Otherwise, Alice and Bob will extract bit keys from new measurements, continue to launch error reconciliation processes until they obtain the same keys. 

\section{Theoretical Analysis}
The existing LSB key extraction schemes have been believed secure, based on the fundamental assumption that wireless signals received at two locations are uncorrelated, when they were separated by more than half a wavelength apart. However, some latest work has observed that this assumption does not hold in some circumstances~\cite{6566763,7321830}. In this section, we provide a formal theoretical verification for the existence of correlation between neighboring links in both indoor and outdoor environments. The important
notations used in our analysis are defined in Table I:

\begin{table}[htbp]
    \caption{\label{tab:regex}Important Notations}
    \centering
    \begin{tabular}{c|p{200pt}}
    	\hline
    	$S_{i}$ & The $i$th scatterer.  \\
    	\hline
    	$\theta_{S}$ & The angle of arrival (AOA) of the wave traveling from the scatterer toward the mobile user. \\
    	\hline
    	$R$ & Radius of scatterer area. \\
        \hline
        $\Omega $ & Received link power.  \\
        \hline
        $f_{S}(S)$ & Probability density function (PDF) of scatterers in an area. \\
        \hline
        $d_{UA}$ & Distance between the mobile user and attack node. \\
        \hline
        $d_{RA}$ & Distance between the legitimate receiver and attack node. \\
        \hline
        $d_{as}$ & Distance between the attack node and the scatterer. \\
        \hline
        $d_{us}$ & Distance between the mobile user and the scatterer. \\
        \hline
        $d_{RA}$ & Distance between the legitimate receiver and the scatterer. \\
        \hline
        $\lambda$ & Wavelength. \\
        \hline

    \end{tabular}
\end{table}

\subsection{Channel Correlation in Outdoor Environment}

\begin{figure}[H]
	\centering
	\includegraphics[scale=0.31]{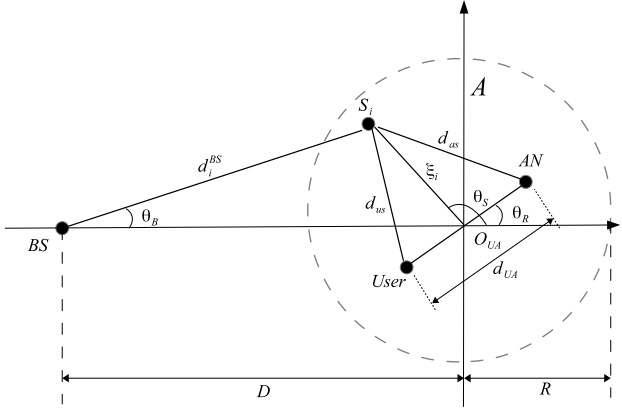}
	\caption{Geometrical configuration of the channel model in outdoor environment.}\label{theoretic1}
\end{figure}

First of all, the wireless communication in an outdoor cellular system is considered,
where a mobile user equipped with one antenna communicates with a base station (BS), as shown in Figure~\ref{theoretic1}. To simplify
the presentation, but without loss of generality, we only consider
the downlink (i.e., BS transmitting to the user) in the following
analysis. Such consideration is representative, as realistic cellular communication is usually dominated by downlink traffic. Moreover, the distance between the BS and the mobile user is much greater than the distance between each antenna element of BS, we will treat the BS as one node in the following discussion. We assume that there is no scatterer around the BS, since the BS antennas are typically installed at a high place. Attack nodes (AN) are placed by the adversary in any potential location. Moreover, we
assume that scatterers are randomly distributed around the mobile user, inside a circular area $A$ around the user's receiving antenna and the attack node, as
illustrated in Figure~\ref{theoretic1}. In Figure~\ref{theoretic1}, the radius of area $A$ is $R$, and the center of $A$ is the midpoint of the user and attack node. We further assume that the number of scatterers is large, and the mobile user and attack node receive the signal from surrounding scatterers, so the line-of-sight (LOS) paths are not considered in our discussion. Our derivation is inspired by the channel correlation analysis in classical MIMO system, first proposed in \cite{837052}.


We start our discussion by defining the correlation coefficient between two neighboring links BS$\to$User and BS$\to$ AN as follows:
\theoremstyle{definition}
\begin{definition}
	The normalized cross correlation coefficient between the neighboring links $BS$ to $User$ and $BS$ to $AN$ is expressed as:
	\begin{equation}
	\rho_{bu,ba} = \frac{E\{h_{bu} \cdot h_{ba}^{*}\}}{\sqrt{\Omega_{bu} \cdot \Omega_{ba}}},
	\end{equation}
	where $*$ is the complex conjugate, $h_{bu}$ and $h_{ba}$ denote the received signal at the user and the attack node, respectively, $\Omega_{bu}$ and $\Omega_{ba}$ are the received link power at the user and the attack node, respectively.
\end{definition}

The received signal $h_{bu}$ between the mobile user and the  BS is given by \cite{995514} (the LOS component is neglected),
\begin{multline}
h_{bu} = \lim_{N \to \infty} \frac{1}{\sqrt{N}} \sum_{i=1}^{N} g_{i} (d_{i}^{BS} \cdot \xi_{i} / D)^{-n/2}\\ 
\cdot exp \{j\psi_{i}-j \frac{2\pi}{\lambda} (d_{i}^{BS}+d_{as})\}.
\end{multline}
where $N$ is the number of scatterers; $g_{i}$ is the amplitude of the wave scattered by the $i$th scatterer; $\psi_{i}$ is the phase shift introduced by the $i$th scatterer, respectively; $d_{i}^{BS}$ and $d_{as}$ are the distances shown in Figure~\ref{theoretic1}; $\lambda$ denotes the wavelength. The term $(d_{i}^{BS} \cdot \xi_{i} / D)^{-n/2}$ accounts for the power loss relative to the distance $D$ between the user and the BS with path loss exponent $n$. The total received power $\Omega_{bu}$ of this link is expressed as

\begin{equation}
\Omega_{bu} = \lim_{N \to \infty} \frac{1}{N} \sum_{i=1}^{N} E\{g_{i}^{2}\}(d_{i}^{BS} \cdot \xi_{i} / D)^{-n}
\end{equation}

We assume all links have equal received power, i.e., $\Omega_{bu}=\Omega_{ba}=\Omega$. By substituting Eq.(2) and Eq.(3) into Eq.(1), the normalized
correlation coefficient between the neighboring links BS$\to$ User and BS$\to$AN can be derived as follows, 
\begin{multline}
\rho_{bu,ba} = \frac{1}{\Omega}\lim_{N \to \infty} \frac{1}{N} \sum_{i=1}^{N} E\{g_{i}^{2}\}(d_{i}^{BS} \cdot \xi_{i} / D)^{-n} \\ \cdot exp \{-j \frac{2\pi}{\lambda}(d_{as}-d_{us})\}.
\end{multline}

In particular, we assume the scatterers be
independently distributed according to some 2-D probability density
function (PDF) $f_{S}(S)$ on the circular area $A$ as shown in Figure~\ref{theoretic1}. When $N$ becomes large, the diffuse power scattered by the $i$th scatterer has quite small contribution out of the total $\Omega$, which is proportional to $E\{g_{i}^{2}\}/N$. This is equal to the infinitesimal power coming from the different area $d_{S}$ with probability $f_{S}(S)$, i.e., $E\{g_{i}^{2}\}/N = f_{S}(S)d_{S}$. Therefore, Eq.(4) can be rewritten in the following integral form:

\begin{multline}
\rho_{bu,ba} = \frac{1}{\Omega} \int_A (d_{i}^{BS} \cdot \xi_{i} / D)^{-n} \\ \cdot exp \{-j \frac{2\pi}{\lambda}(d_{as}-d_{us})\}f_{S}(S)dS,
\end{multline}

Since we assume scatterers are randomly distributed inside the circular area $A$ around the user and attack node, Eq.(5) can be written as
\begin{multline}
\rho_{bu,ba} = \frac{1}{\Omega} \int_0^R \int_{-\pi}^{\pi} (d_{i}^{BS} \cdot \xi_{i} / D)^{-n} \cdot exp \{-j \frac{2\pi}{\lambda} \\ \cdot (d_{as}-d_{us})\}f(\theta_{S}, \xi_{i}) \,d(\theta_{S})d(\xi_{i}).
\end{multline}
where the $f(\theta_{S}, \xi_{i})$ is the PDF of the locations of scatterers relative to the user and attack node with $\theta_{S}$ and distance $\xi_{i}$ and $d_{i}^{BS}$, $\theta_{S}$ is the angle of arrival (AOA) of the wave traveling from the scatterer toward the mobile user.

According to the laws of cosine and sine in \cite{4273749}, we get
\begin{equation}
\begin{split}
&d_{as}^{2} = d_{UA}^{2}/4+\xi_{i}^2-d_{UA} \cdot \xi_{i} \cdot cos(\theta_{S}-\theta_{R}) \\
&d_{us}^{2} = d_{UA}^{2}/4+\xi_{i}^2+d_{UA} \cdot \xi_{i} \cdot cos(\theta_{S}-\theta_{R})
\end{split}
\end{equation}
and
\begin{equation}
\frac{D}{sin(\theta_{S}-\theta_{B})} = \frac{\xi_{i}}{sin(\theta_{B})} = \frac{d_{i}^{BS}}{sin(\theta_{S})},
\end{equation}
where the $d_{UA}$ is the distance between the mobile user and the attack node.

In the outdoor environment, the assumption of $D \gg R \gg d_{UA}$ is realistic. Therefore, the difference of path lengths can be approximated as
\begin{equation}
\begin{split}
-(d_{as}-d_{us}) & \approx d_{UA} \cdot cos(\theta_{S}-\theta_{R}) \\
d_{i}^{BS} & \approx D
\end{split}
\end{equation}

Substituting the arguments in Eq.(6) with Eq.(7)(8)(9) yields
\begin{multline}
\rho_{bu,ba} = \frac{1}{\Omega} \int_0^R \int_{-\pi}^{\pi} (\xi_{i})^{-n}  exp \{-j \frac{2\pi}{\lambda} \\ \cdot d_{UA} \cdot cos(\theta_{S}-\theta_{R})\}f(\theta_{S}, \xi_{i}) \,d(\theta_{S})d(\xi_{i}).
\end{multline}
Since the scatterers are uniformly distributed inside the circular ring, we can use the PDF as $f(\theta_{S}, \xi_{i}) = 1 / 2\pi R$.

Then, we obtain
\begin{multline}
\rho_{bu,ba} = \frac{1}{\Omega} \int_0^R \int_{-\pi}^{\pi} (\xi_{i})^{-n}  exp \{-j \frac{2\pi}{\lambda} \\ \cdot d_{UA} \cdot cos(\theta_{S}-\theta_{R})\} \frac{1}{2\pi R} \,d(\theta_{S})d(\xi_{i}) \\
= \frac{1}{\Omega} \int_0^R \int_{-\pi}^{\pi} \frac{(\xi_{i})^{-n}} {2\pi R} exp \{-j2\pi\frac{d_{UA}}{\lambda} \\ *cos(\theta_{S}-\theta_{R})\} \,d(\theta_{S})d(\xi_{i})
\end{multline}

\begin{figure}[H]
    \centering
    \includegraphics[scale=0.5]{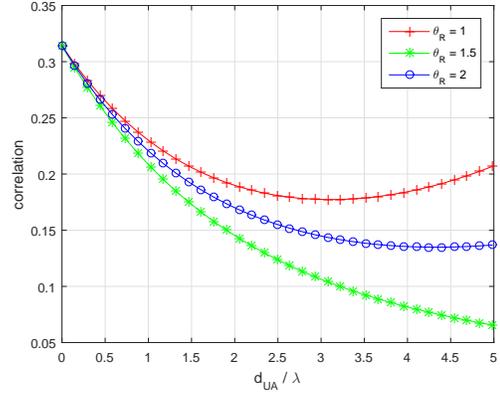}
    \caption{Channel correlation coefficient $\rho_{bu,ba}$ versus distance ratio $d_{UA}/\lambda$.}\label{outdoor}
\end{figure}

We plot the correlation coefficient $\rho_{bu,ba}$ as a function of the ratio $d_{UA}/\lambda$ in Figure~\ref{outdoor}, in which we assume the distance $D$ is fixed. From this numerical result, it can be observed that there exists correlation, even if the distance $d_{UA}$ between the mobile user and the attack node is greater than half wavelength. For instance, $\rho_{bu,ba} = 0.21$ when $d_{UA}$ is equal to $5$ wavelength, and $\theta_{R}=1$. Moreover, the angle $\theta_{R}$ affects the correlation coefficient $\rho_{tr,ta}$ dramatically in this model.

\subsection{Channel Correlation in Indoor Environment}
\begin{figure}[H]
    \centering
    \includegraphics[scale=0.31]{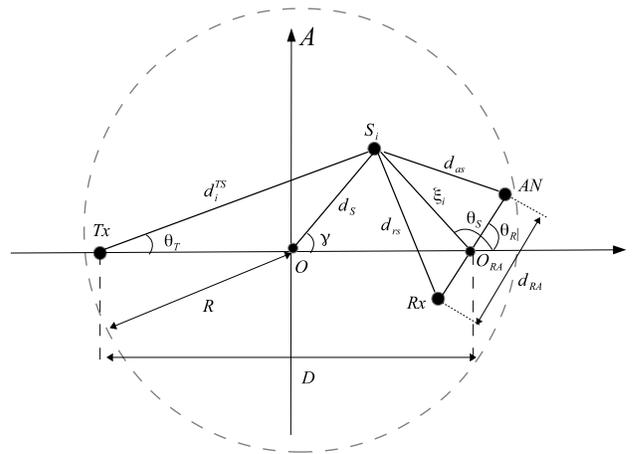}
    \caption{Geometrical configuration of the channel model in indoor environment.}\label{theoretic2}
\end{figure}

In indoor environments, two communication ends, the transmitter (Tx) and the receiver (Rx), generally are not far from each other and surrounded by scatterers nearby. In this case, we build a model, in which a big scatterer-ring area $A$ encloses both the transmitter and the receivers (including attack nodes), as depicted in Figure~\ref{theoretic2}. In this circular area $A$, the radius is $R$, and the center $O$ is the midpoint of the Tx and $O_{RA}$, where $O_{RA}$ is the midpoint of the Rx and the attack node (AN). In Figure~\ref{theoretic2}, we follow most of the notations as those defined in Figure~\ref{theoretic1}. Comparing to the outdoor model as shown in Figure~\ref{theoretic1}, in the indoor model, we substitute the notations ``Tx" and ``Rx" for ``BS" and ``User", respectively. Moreover, to facilitate the derivation, we substitute $d_{i}^{TS}$, $d_{RA}$, $d_{rs}$ and $\theta_{T}$ for $d_{i}^{BS}$, $d_{UA}$, $d_{us}$ and $\theta_{B}$, respectively, and introduce the new variable $\gamma$ to denote the angle of the scatterer in the polar coordinate system, as illustrated in Figure~\ref{theoretic2}.

In Figure~\ref{theoretic2}, the correlation coefficient between two neighboring links  Tx$\to$Rx and Tx$\to$AN is defined as follows:
\theoremstyle{definition}
\begin{definition}
	The normalized cross correlation coefficient between the neighboring links $Tx$ to $Rx$ and $Tx$ to $AN$ is expressed as:
	\begin{equation}
		\rho_{tr,ta} = \frac{E\{h_{tr} \cdot h_{ta}^{*}\}}{\sqrt{\Omega_{tr} \cdot \Omega_{ta}}},
	\end{equation}
	where $*$ is the complex conjugate, $h_{tr}$ and $h_{ta}$ denote the received signal at the receiver and the attack node, respectively, $\Omega_{tr}$ and $\Omega_{ta}$ are the received link power at the receiver and the attack node, respectively
\end{definition}

To discuss the correlation between the neighboring links Tx$\to$Rx and Tx$\to$AN in this model, we still use Eq.(5) to yield the correlation coefficient function. The Eq.(5) is rearranged in the indoor model as follows:
\begin{multline}
\rho_{tr,ta} = \frac{1}{\Omega} \int_A (d_{i}^{TS} \cdot \xi_{i} / D)^{-n} \\ \cdot exp \{-j \frac{2\pi}{\lambda}(d_{as}-d_{rs})\}f_{S}(S)dS,
\end{multline}
where $\Omega$ is the received link power (equal power
for all radio links is assumed).

According to the laws of cosine and sine in \cite{4273749}, we have
\begin{equation}
\begin{split}
(d_{i}^{TS})^{2} &= d_{S}^{2}+D^2/4+D \cdot d_{S}cos(\gamma) \\
\xi_{i}^{2} &= d_{S}^{2}+D^2/4-D \cdot d_{S}cos(\gamma)
\end{split}
\end{equation}
and
\begin{equation}
\begin{split}
sin(\theta_{T}) &= d_{S}sin(\gamma)/d_{i}^{TS} \\
&= d_{S}sin(\gamma)/ \sqrt{d_{S}^{2}+D^2/4+D \cdot d_{S}cos(\gamma)}
\end{split}
\end{equation}

\begin{equation}
\begin{split}
cos(\theta_{T}) &= (d_{S}cos(\gamma)+D/2) / d_{i}^{TS} \\
&= (d_{S}cos(\gamma)+D/2)/ \sqrt{d_{S}^{2}+D^2/4+D \cdot d_{S}cos(\gamma)}
\end{split}
\end{equation}

\begin{equation}
\begin{split}
sin(\theta_{S}) &= d_{S}sin(\gamma) / \xi_{i} \\
&= d_{S}sin(\gamma)/ \sqrt{d_{S}^{2}+D^2/4-R \cdot d_{S}cos(\gamma)}
\end{split}
\end{equation}

\begin{equation}
\begin{split}
cos(\theta_{S}) &= (d_{S}cos(\gamma)-D/2) / \xi_{i} \\
&= (d_{S}cos(\gamma)-D/2)/ \sqrt{d_{S}^{2}+D^2/4-R \cdot d_{S}cos(\gamma)}
\end{split}
\end{equation}

By doing the substitution and rearrangements, we obtain the approximation of $-(d_{as}-d_{rs})$ as
\begin{equation}
-(d_{as}-d_{rs}) \approx \frac{d_{RA} \cdot (d_{S}cos(\gamma-\theta_{R})-D/2 \cdot cos(\theta_{R}))}{\sqrt{d_{S}^{2}+D^2/4-D \cdot d_{S}cos(\gamma)}}
\end{equation}

Since the scatterers are uniformly distributed inside the circular ring, we can use the PDF as $f_{S}(S)=f(d_{S}, \gamma) = \frac{1}{2\pi R}$. Substituting the arguments in Eq.(13) yields the correlation coefficient $\rho_{tr,ta}$ between neighboring links Tx$\to$Rx and Tx$\to$AN as
\begin{multline}
\rho_{tr,ta} = \frac{D^{n}}{\Omega} \int_0^R \int_{-\pi}^{\pi}((d_{S}^2+D^2/4)^2-D^2d_{S}^2cos^2\gamma)^{-n/2} \\ \frac{1}{2\pi R} \cdot exp\{\frac{-j2\pi \cdot  d_{RA}(d_{S}cos(\gamma-\theta_{R})-D/2 \cdot cos(\theta_{R}))}{\lambda\sqrt{d_{S}^{2}+D^2/4-D \cdot d_{S}cos(\gamma)}}\} \\ d_{\gamma}d(d_{S})
\end{multline}

\begin{figure}[H]
    \centering
    \includegraphics[scale=0.5]{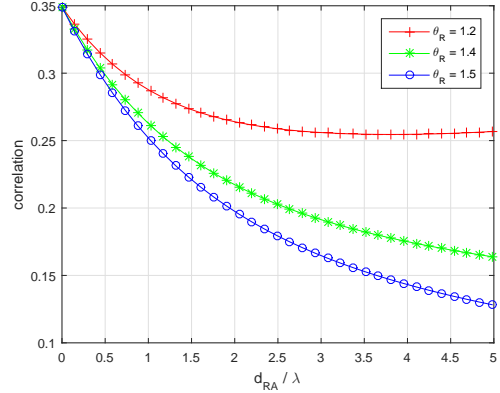}
    \caption{Channel correlation coefficient $\rho_{tr,ta}$ versus distance ratio $d_{RA}/\lambda$.}\label{indoor}
\end{figure}

The correlation coefficient $\rho_{tr,ta}$ as a function of the ratio $d_{RA}/\lambda$ is plotted in Figure~\ref{indoor}. It can be observed that there still exists correlation, even if the distance $d_{RA}$ between the legitimate receiver and the attack node is greater than half wavelength. For instance, the correlation coefficient $\rho_{tr,ta} = 0.26$ when $d_{RA}$ is equal to $5$ wavelength, and $\theta_{R}=1.2$. In contrast to the outdoor model, the correlation is a decreasing function of the angle $\theta_{R}$, for a certain distance ratio $d_{RA}/\lambda$. Moreover, comparing to the outdoor model, the correlation coefficient $\rho_{tr,ta}$ is more sensitive to $\theta_{R}$. 

In conclusion, the wiretap channel (i.e., Tx$\to$AN) and the legitimate channel (i.e., Tx$\to$Rx) still have relatively strong correlation, even if the attack node is far away from the transmitter and receiver (here the ``far away" means that the distance between the attack node and receiver is greater than $\lambda/2$). Therefore, the adversary can infer the secrect key bits based on his measurements of channel states, which makes existing key extraction system vulnerable. In the following section, we propose a class of attacks, called statistical inference attacks (SIAs), to reveal the vulnerability of LSB key extraction system.

\section{Statistical Inference Attack}\label{sec:attack}
\begin{figure}[htbp]
    \centering
    \subfloat[Case I: Using links of different tx and rx locations.]{\includegraphics[scale=0.09]{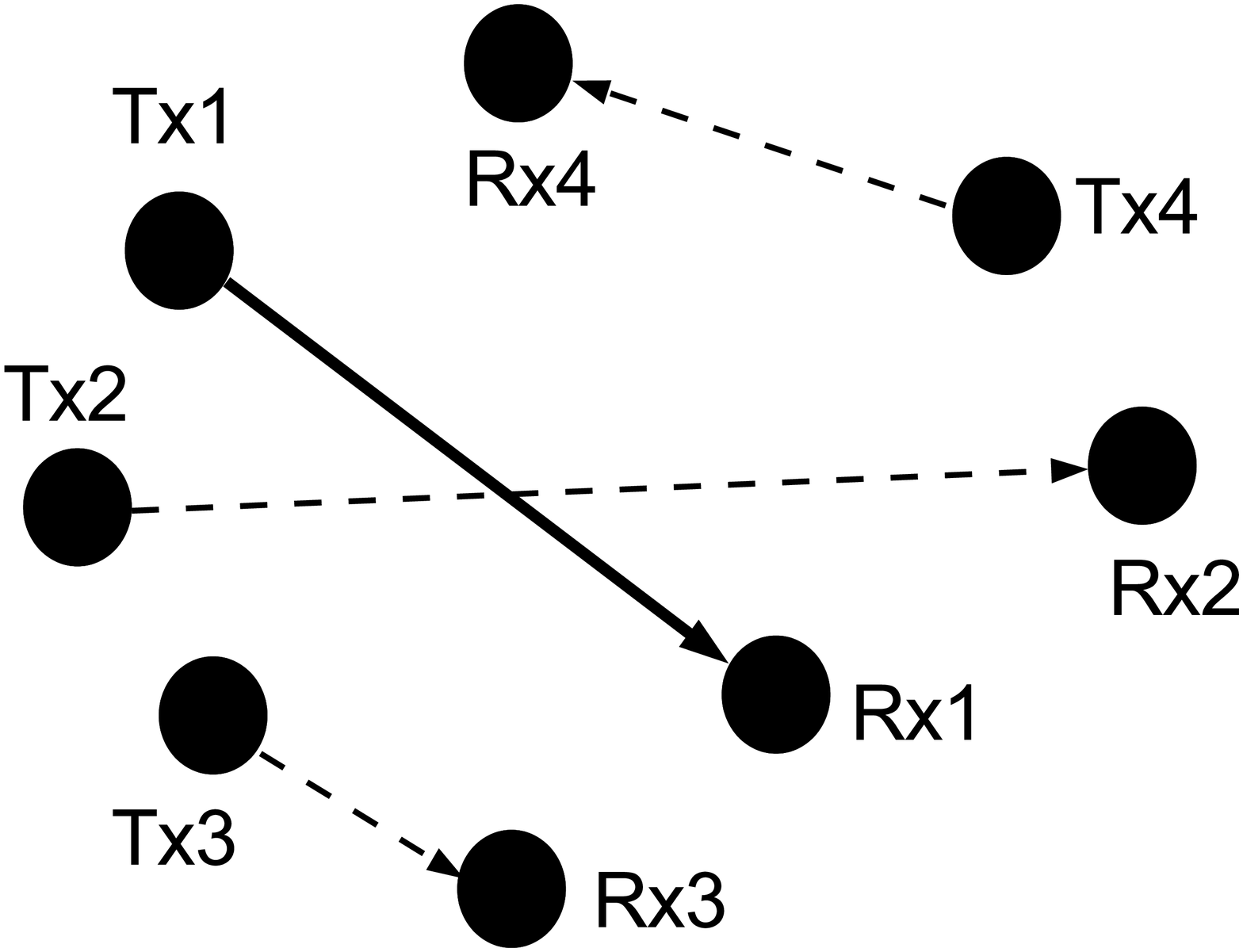}
        }
    \hfil
    \subfloat[Case II: Using links of the same tx locations.]{\includegraphics[scale=0.09]{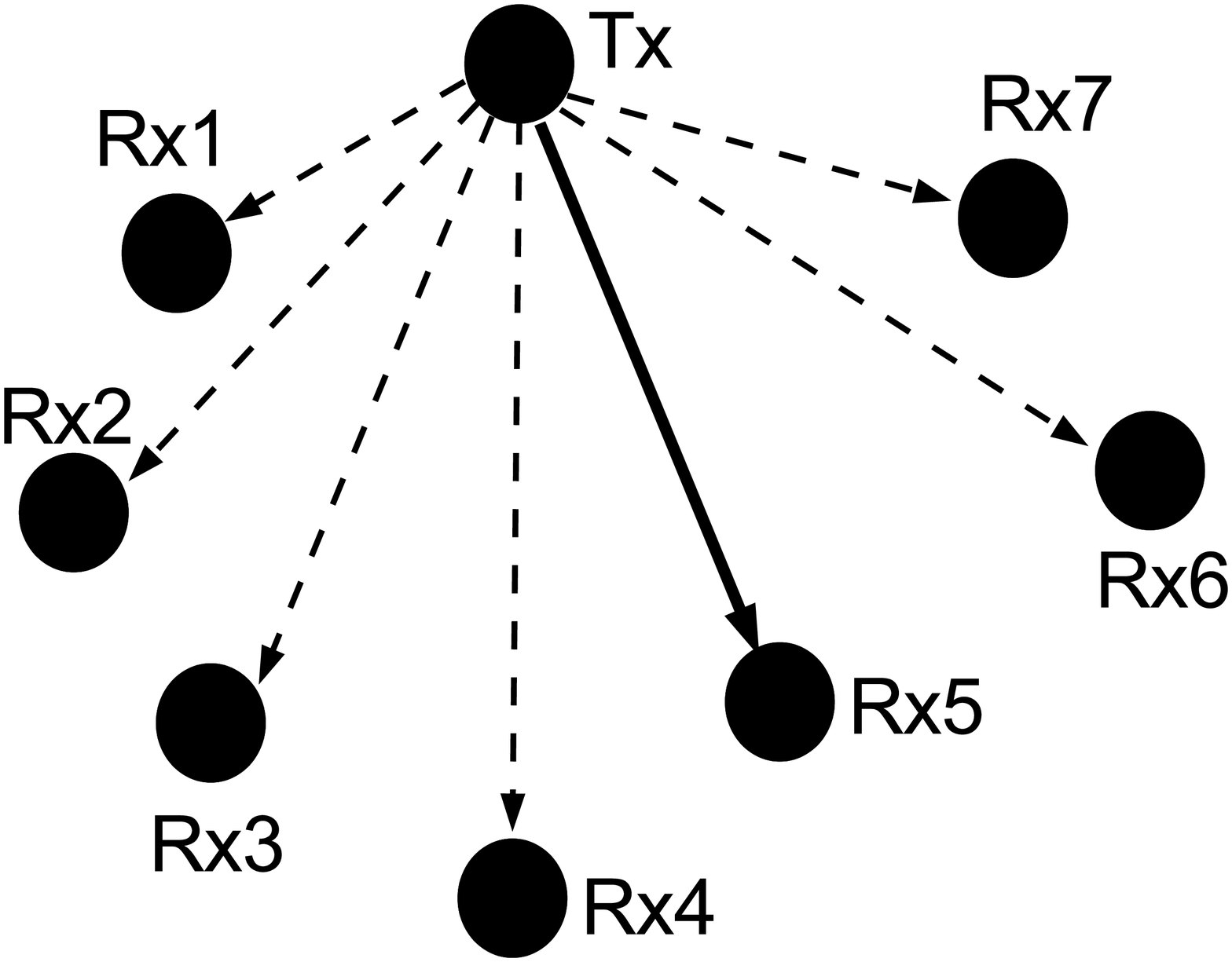}
        }
    \hfil
    \subfloat[Case III: Using signatures of the same link, but measured at different times.]{\includegraphics[scale=0.09]{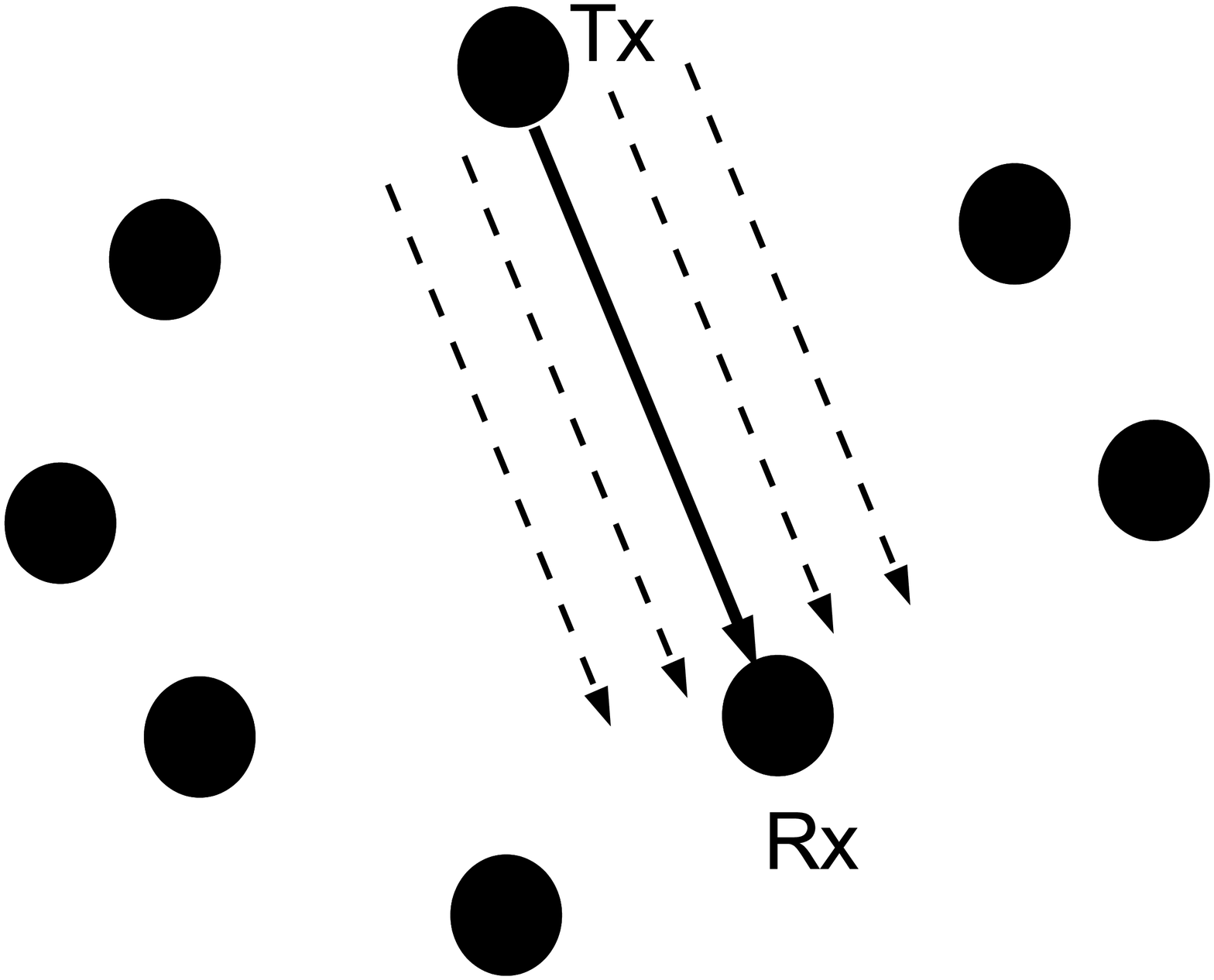}
        }
    \caption{Attacks in three scenarios, solid lines denote the target links, dotted lines denote the existing links.}\label{attack}
\end{figure}

In the previous section, we have theoretically verified the existence of correlation between neighboring links. The remaining issue is how this correlation may be exploited to infer secret keys in practice. In this section, we apply several ML-based algorithms to propose statistical inference attacks (SIAs) to infer keys by exploiting this correlation.
Depending on the information available for the training of the ML
model, we consider the following three scenarios for SIAs:
\begin{itemize}
    \item [(a)] Inference based on links disjoint from the target link (i.e., links of different transmitters and
    receivers from the target link)
    \item [(b)] Inference based on links sharing the same transmitter as the target link
    \item [(c)] Inference based on historical signatures of the target link
\end{itemize}
These scenarios are illustrated in Figure~\ref{attack},
respectively, and are elaborated as follows.

\subsection{SIA based on disjoint links}
This case is pertinent to the scenario where the adversary has
pre-knowledge regarding the area in which the target (i.e., the
transmitter and receiver of the target link) will appear, but does
not know the exact location of the target until the target appears.
In this case, the adversary may perform a site survey in the area
before the target appears. During the site survey, the adversary may
collect sample signatures of links throughout the area, and then use
these samples to train a model that represents the signature of an
arbitrary link in the area as a function of the transmitter and
receiver locations of the link. Later in the online inference phase,
the adversary can observe the location of the target, and supply this
information to the trained model to infer the signature of the
target link. SIA to links in a mobile ad hoc network is a typical
example of this scenario.

The topology of ML model for the above link signature inference
is as follows:

\begin{itemize}
    \item [] Input: $S(L_{T}(i), L_{S}(i))$, $L_{T}(i)$, $L_{S}(i)$,
    \item [] Output: $S(L_{T}(t),L_{S}(t))$,
\end{itemize}
where
\begin{itemize}
    \item  $i$ = the index of surveyed links,
    \item  $t$ = the index of the target link,
    \item  $L_{T}(i)$ = the locations of transmitters,
    \item  $L_{S}(i)$ = the locations of receivers,
    \item  $S(L_{T}(i), L_{S}(i))$ = link signatures on the links ($L_{T}(i)$, $L_{S}(i)$),
    \item  $S(L_{T}(t), L_{S}(t))$ = link signature on the target link.
\end{itemize}

\subsection{SIA based on links sharing the same transmitter}
This case applies to the scenario where the adversary does not know
the exact location of the target until the target appears, but has
pre-knowledge on the area in which the target will appear, and the
communication in this area is through a centralized access point
such as a base station or an AP. Typical examples of such scenario
include cellular network and WLAN. In this case, the adversary can
also survey the area before the target appears, during which he collects sample signatures of various downlinks throughout
the area. These sample signatures are then used to train a downlink
signature model of the area, which represents the signature of an
arbitrary as a function of the receiver's location. In the online
inference phase, the adversary observes the location of the target
(the receiver), and supply this information to the trained model to
infer the signature of the target downlink.

The topology of ML model in this case is given as follows:

\begin{itemize}
    \item [] Input: $S(L_{S}(i))$, $L_{S}(i)$,
    \item [] Output: $S(L_{S}(t))$,
\end{itemize}
where
\begin{itemize}
    \item  $i$ = the index of surveyed downlinks,
    \item  $t$ = the index of the target downlink,
    \item  $L_{S}(i)$ = the receiver location of the $i$th surveyed
    downlink,
    \item  $S(L_{S}(i))$ = link signature of the $i$th surveyed downlink,
    \item  $S(L_{S}(t))$ = link signature of the target downlink.
\end{itemize}

\subsection{SIA based on historical signatures of the target link}
This case applies to the scenario where the adversary has
pre-knowledge on the exact location of the target. Such information
can be obtained by the adversary by peeking into the location privacy
of the target. This is especially true if the target's activity or
schedule follows a regular rule.

To infer the signature of the target link at a given time, the adversary may first measure the signatures of the target link at
different times. The sample signatures are then used to train a
model that represents the link signature as a function of time. In
the online inference phase, the adversary simply feeds the desired
time into the trained model to make inference on the signature of
the target link at that time.

The topology of ML model for this case is given as follows:

\begin{itemize}
    \item [] Input: $S(t_{i})$, $t_{i}$,
    \item [] Output: $S(t_{c})$,
\end{itemize}
where
\begin{itemize}
    \item  $i$ = the index of time,
    \item  $t_{i}$ = time $t_{i}$,
    \item  $S(t_{i})$ = link signature at time $t_{i}$,
    \item  $S(t_{c})$ = link signature at the target moment.
\end{itemize}

\section{Experiment Verification}\label{sec:results}
In this section, we evaluate the effectiveness of the above
statistical inference attacks based on the CRAWDAD dataset\cite{1212671}.

\subsection{The CRAWDAD dataset}
In CRAWDAD dataset, over 9300 link signatures are recorded in a
44-node wireless network, which are measured in an indoor
environment with obstacles and scatters. By moving the transmitter
and receiver between node locations 1 - 44, it gives the
number of transmitter and receiver permutations counted as $44*43=1892$. At each permutation of transmitter and receiver, 5
link signatures are measured over a period of about 30 seconds. In this dataset, each link signature is recorded as a 50-component vector.

\begin{figure*}[tbp]
    \centering
    \subfloat[Case I]{\includegraphics[scale=0.39]{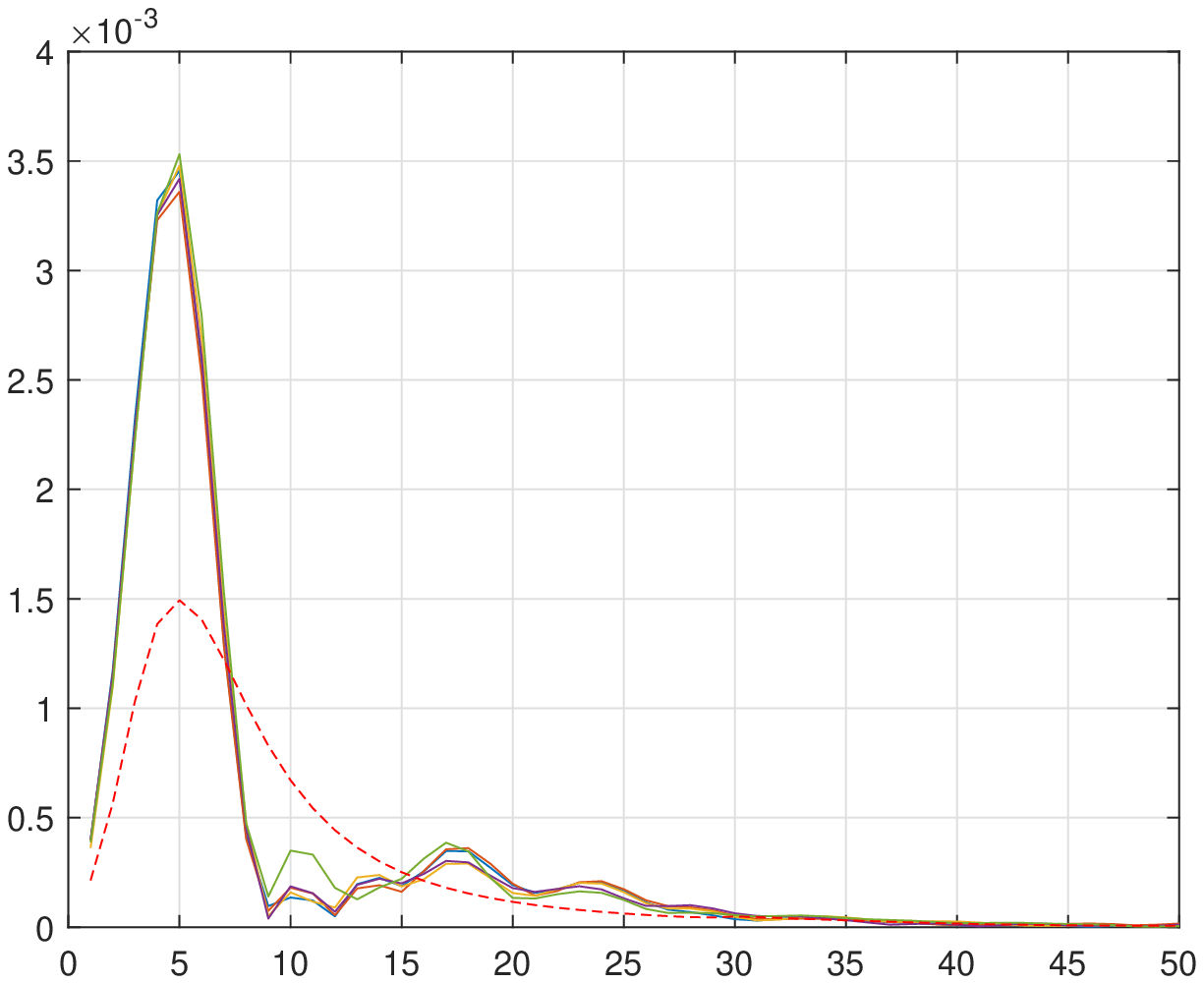}
        \label{fig_first_case}}
    \hfil
    \subfloat[Case II]{\includegraphics[scale=0.39]{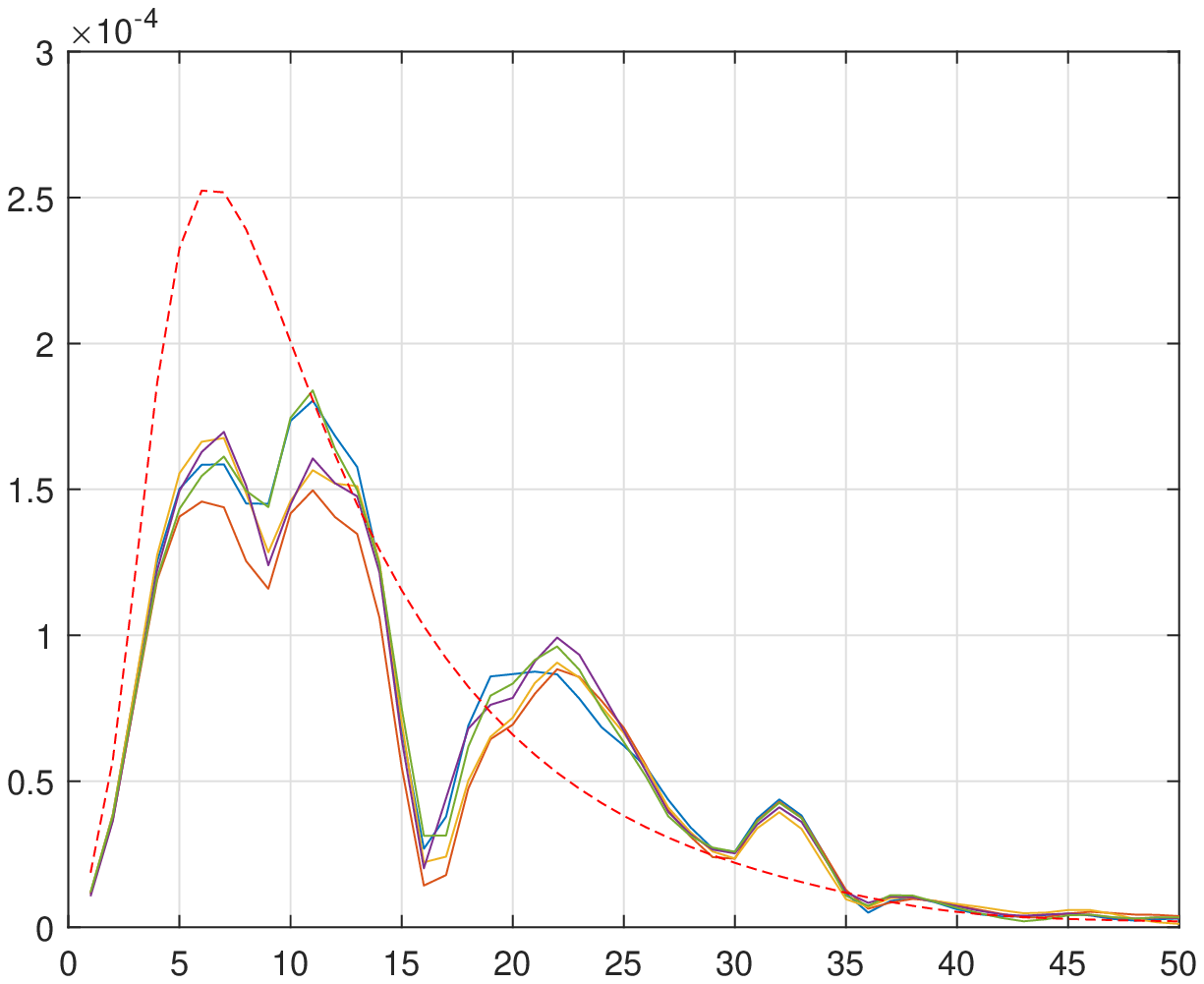}
        \label{fig_second_case}}
    \hfil
    \subfloat[Case III]{\includegraphics[scale=0.39]{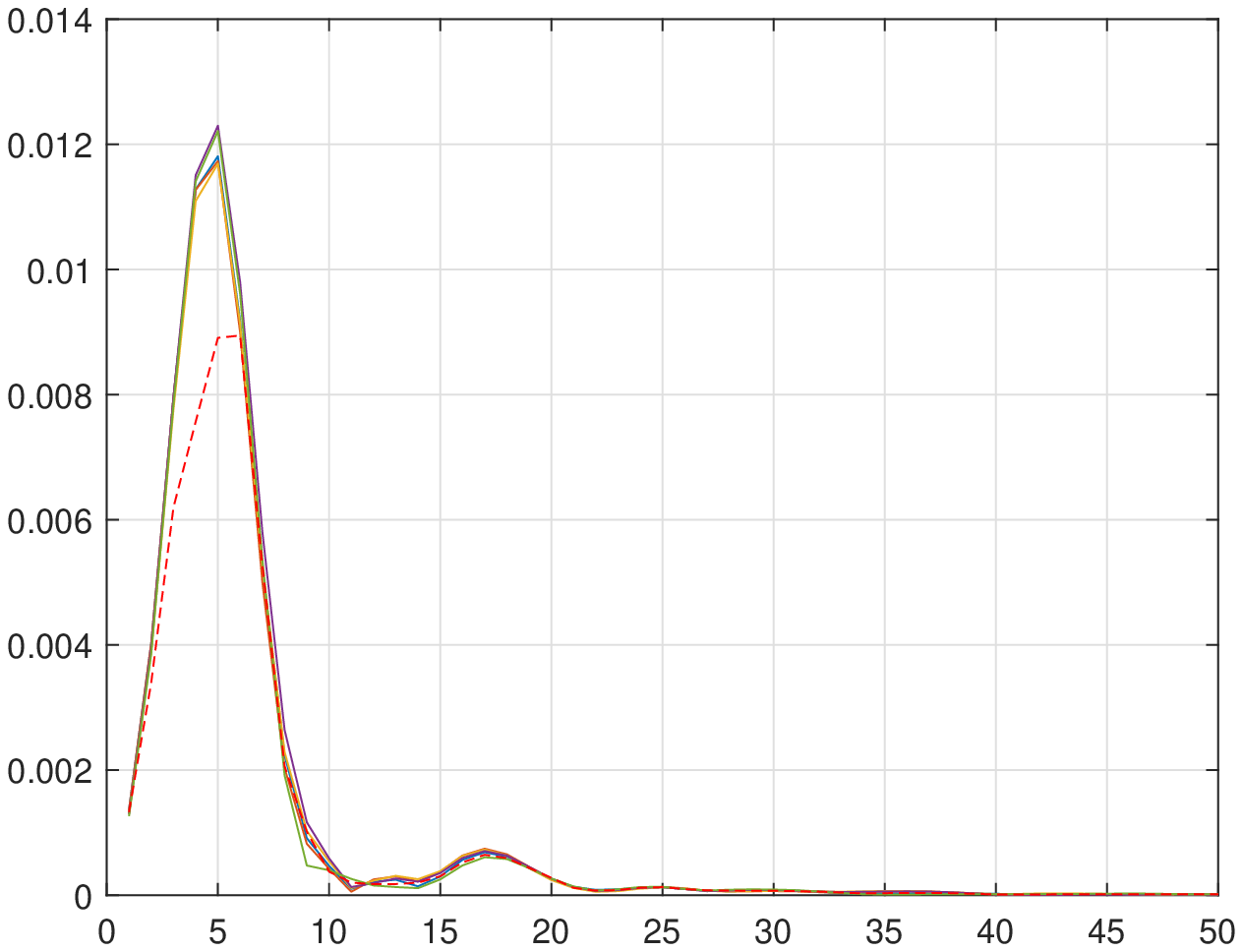}
        \label{fig_third_case}}
    \caption{Link signature inference accuracy. Note that x-axis denotes index of points, y-axis denotes the amplitude; the solid lines represent the true link signatures, and dotted line represents our inferred link signature.}
    \label{fig_sim}
\end{figure*}

\subsection{SIA Results and Analysis}
We first evaluate how accurate the proposed ANNs can infer the
signature of a target link. To this end, we randomly pick a link (a
transmitter-receiver pair) from CRAWDAD as the target link. The five
signatures of the target link are used as ground truth for testing.
Training data are selected from the remaining links in CRAWDAD in
the following way. For case-I SIA, we use all remaining links, in
total $9300-43*2*5=8870$ signatures, as training data. For case-II
SIA, we use all the 43 links that share the same transmitter of the
target link but have a different receiver as the training data. So
in total $43*5=215$ link signatures are included in the training
dataset for case II. For case III SIA, we
randomly pick 4 signatures of the target link and use them as
training data, and the remaining link signature is used for testing.
In a nutshell, in this experiment we use all relevant data for
training to avoid the complicated issue of training data selection.
As a baseline of the performance, our goal here is to see how well
the ANN can do without discriminating the available
training data. The optimization of the inference, e.g., through
training data filtering, is studied shortly.

Different target links were inferred in our SIA experiments. Figure~\ref{fig_sim} plots a typical case for the comparison between the inferred
signature and the ground truth version for the target link $1 \to 4$
under the three SIA cases, respectively. The inferences for other
target links present similar trends, and thus are omitted here due
to space limit. Three observations can be made on Figure~\ref{fig_sim}. First,
in all three cases there exists significant similarity between the
inferred signature and the ground truth, and the trends of curves
match quite well. This observation implies that there are indeed
correlations between neighboring wireless links, even when their
separation is farther than half a wavelength, and these correlations
are harnessed by ANN models in the experiments for inference.
Second, the inferred signature presents different accuracy in the
three cases. This is not surprising, because the inferences are
based on different amount of knowledge about the target. In
particular, training data in case III is the closest to what is being
inferred, and therefore the inference accuracy in that case is the
highest among the three cases. Third, the inferred signature is much
smoother than the truth version. The current ANN models cannot
capture enough high frequency details in the correlation to make a
better inference. This observation suggests that the ANN models we
are using may not be the optimal ones and there are sufficient rooms
for improvement from a ML's point of view.

\begin{figure}[H]
	\centering
	\includegraphics[scale=0.5]{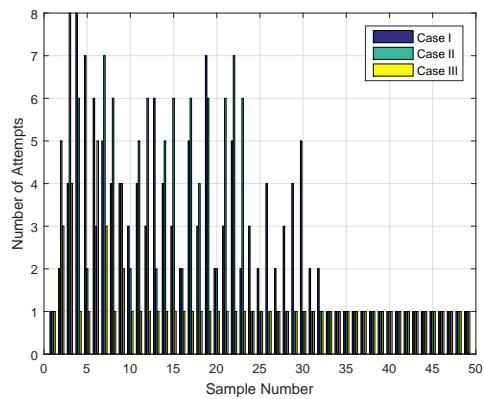}
	\caption{Average number of guesses needed to find the 50 points in a link signature (ANN).}\vspace{1mm}\label{guess}
\end{figure}

We study the impact of inference accuracy on the security strength
of LSB key extraction as shown in Figure~\ref{guess}. In this experiment, the goal of
the adversary is to figure out, or guess, the secret key extracted
from the true link signature, from the inferred version of the
signature. In particular, we assume that each time-series point on
the true link signature is represented as a 5-bit binary number
according to the quantization scheme described in Section III.B
(32-interval quantization). So in total a 250-bit binary string can be
extracted from the 50-point true link signature. We pick a 75-bit string as the true key in our experiment. To guess the true key, the adversary uses trial and error, starting from the 75 bits
quantization of the inferred signature (5 bits per point, $15*5$-bit
binaries in total). In each round of trial and error, the adversary
explores the key search space by incrementing or decrementing by one
to one of the $15*5$-bit binaries, where the exploration is sequential
over the $15*5$-bit binaries. Under such an inference attack, the
security strength of the LSB key extraction can be measured by the
average number of trials needed to find the true key. Equivalently,
this metric can be normalized on a per-point basis, i.e., measured
by the average number of guesses required to find the true 5-bit
quantization for a point on the link signature.

Figure~\ref{guess} plots the average required number of guesses for each of
the whole 50 points on a link signature under the three SIA cases. The
average is based on 26 target links randomly selected from CRAWDAD.
From this figure, it can be observed that for cases I and II on
average at most 8 guesses are enough to find the true quantization
of a point using the ANNs. In contrast, to find a 5-bit
quantization, a brutal force search algorithm needs on average 16
guesses. Therefore, for the 75-bit key in this experiment, the key search
space of the brutal force algorithm is $16^{15}$, or $2^{60}$.
Using the proposed ANNs, the key search space is at most $8^{15}$,
or $2^{45}$: a reduction of $2^{15}$ compared to the brutal force
search! On a computer with 4-core Intel CPU (2.0 GHz CPU clock speed), it takes about 4,000 hours to find the correct secret key by brutal force algorithm, however, it takes only about 1 hour and 13 minutes to find it by our proposed mechanism. Note that this is the upper bound (worst case) key search
space for the case I and case II SIAs, because the number of required
guesses per point is much smaller than 8 for the points on the tail
of the link signature. For example, only one guess is needed for
points after index 33. Furthermore, it can also be observed that SIA
in case III is able to figure out the true key much more efficiently,
as $90\%$ points can be found in just one guess in that case.

\begin{figure}[H]
	\centering
	\includegraphics[scale=0.5]{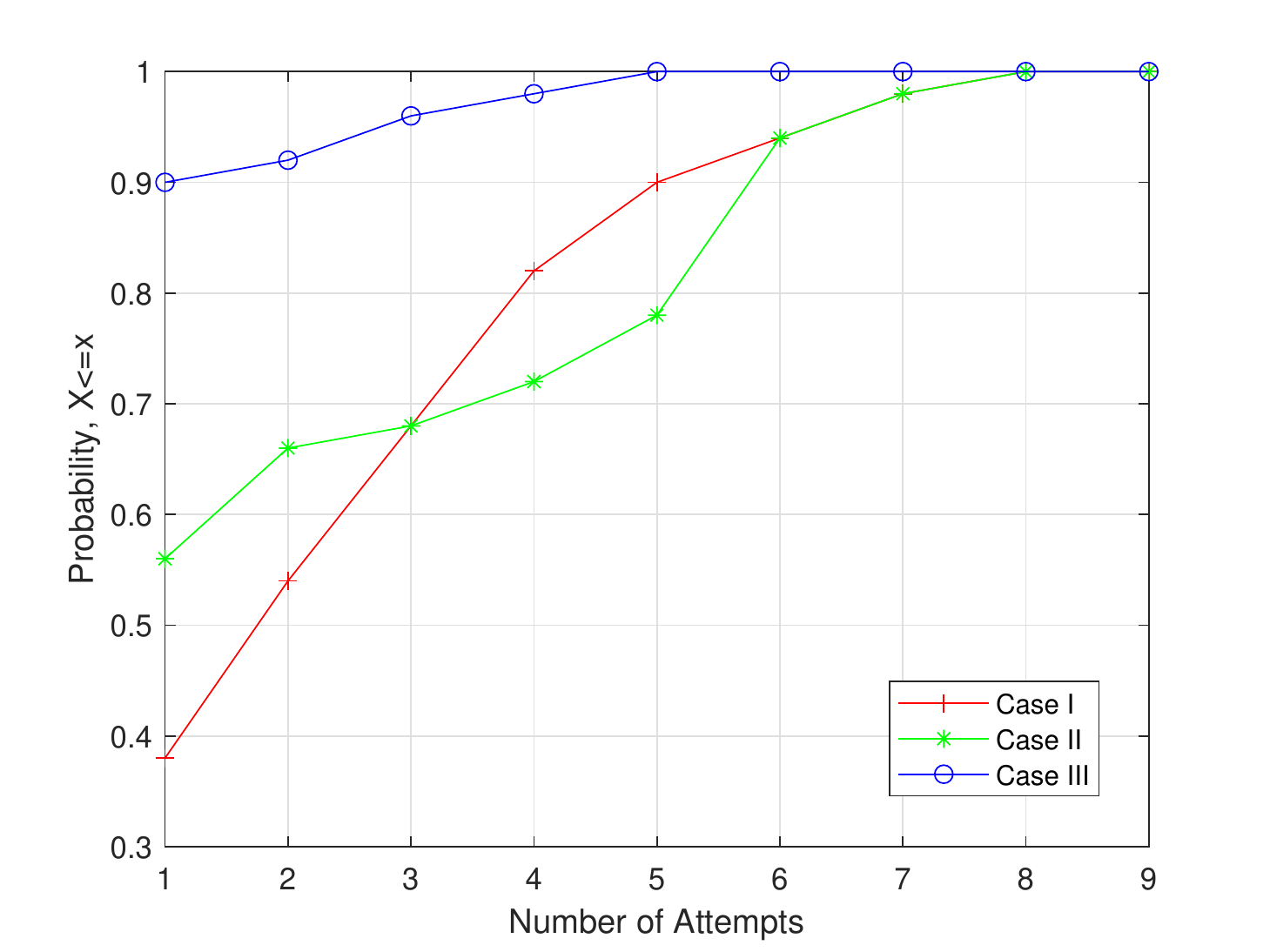}
	\caption{Comparison of inference performance in different cases by using ANN.}\vspace{1mm}\label{caseann}
\end{figure}

To obtain a statistical view about the strength of the proposed
SIAs, Figure~\ref{caseann} compares the CDF (cumulative distribution function)
of the number of guesses needed per point under various SIA cases.
The CDFs in the figure are calculated based on the same 26 target
links as in Figure~\ref{guess}. This figure shows that statistically the
relative strength of SIAs are case III $>$ case II $>$ case I. For
example, case III can find $90\%$ points in just one round, while
$56\%$ points are found in one round in case II, and only $38\%$ are
found in one round in case I. This trend is aligned with the
inference accuracies as observed in Figure~\ref{fig_sim}.

\begin{figure}[H]
	\centering
	\subfloat[Case I]{\includegraphics[scale=0.5]{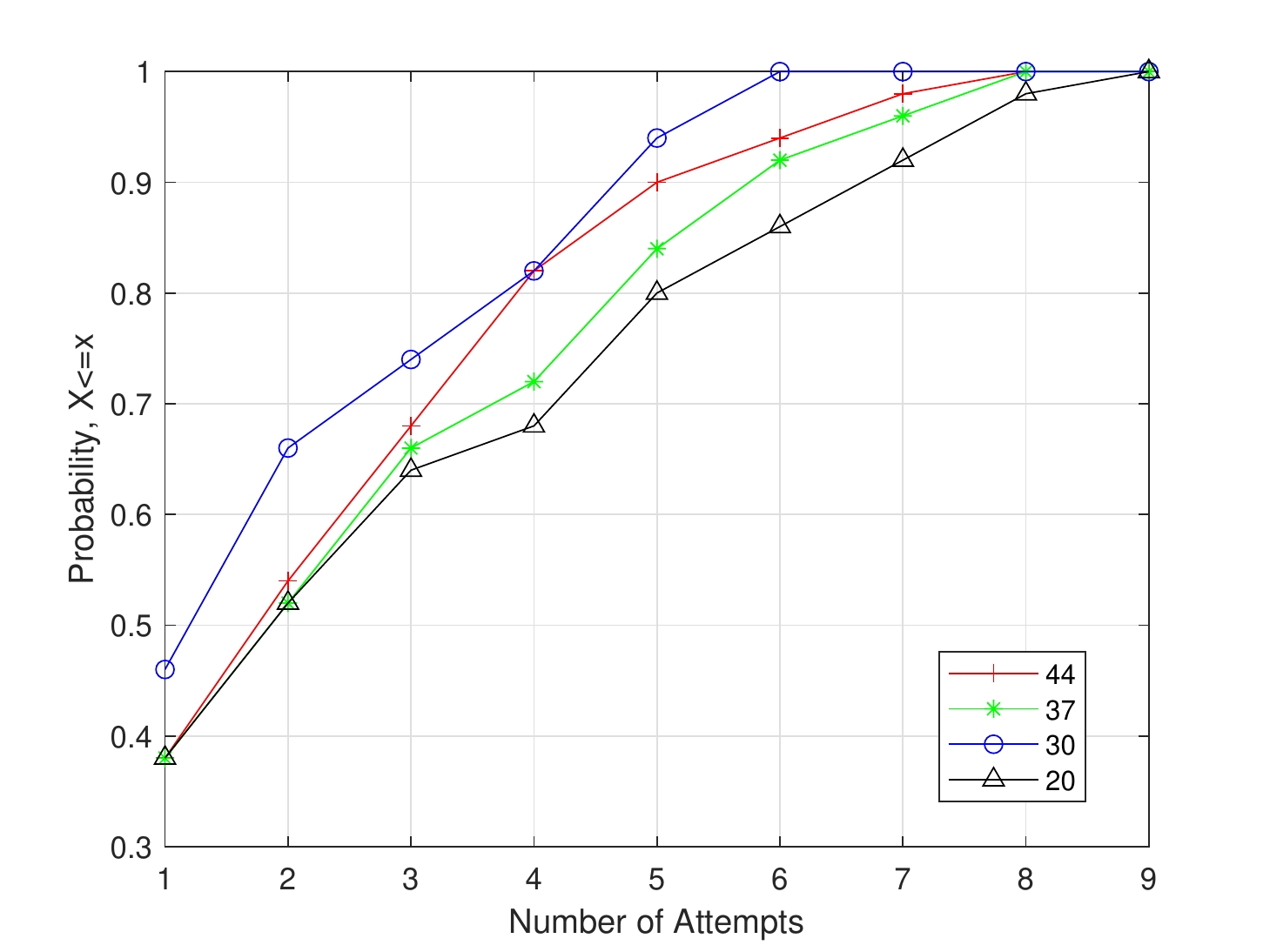}
	}
	\hfil
	\subfloat[Case II]{\includegraphics[scale=0.5]{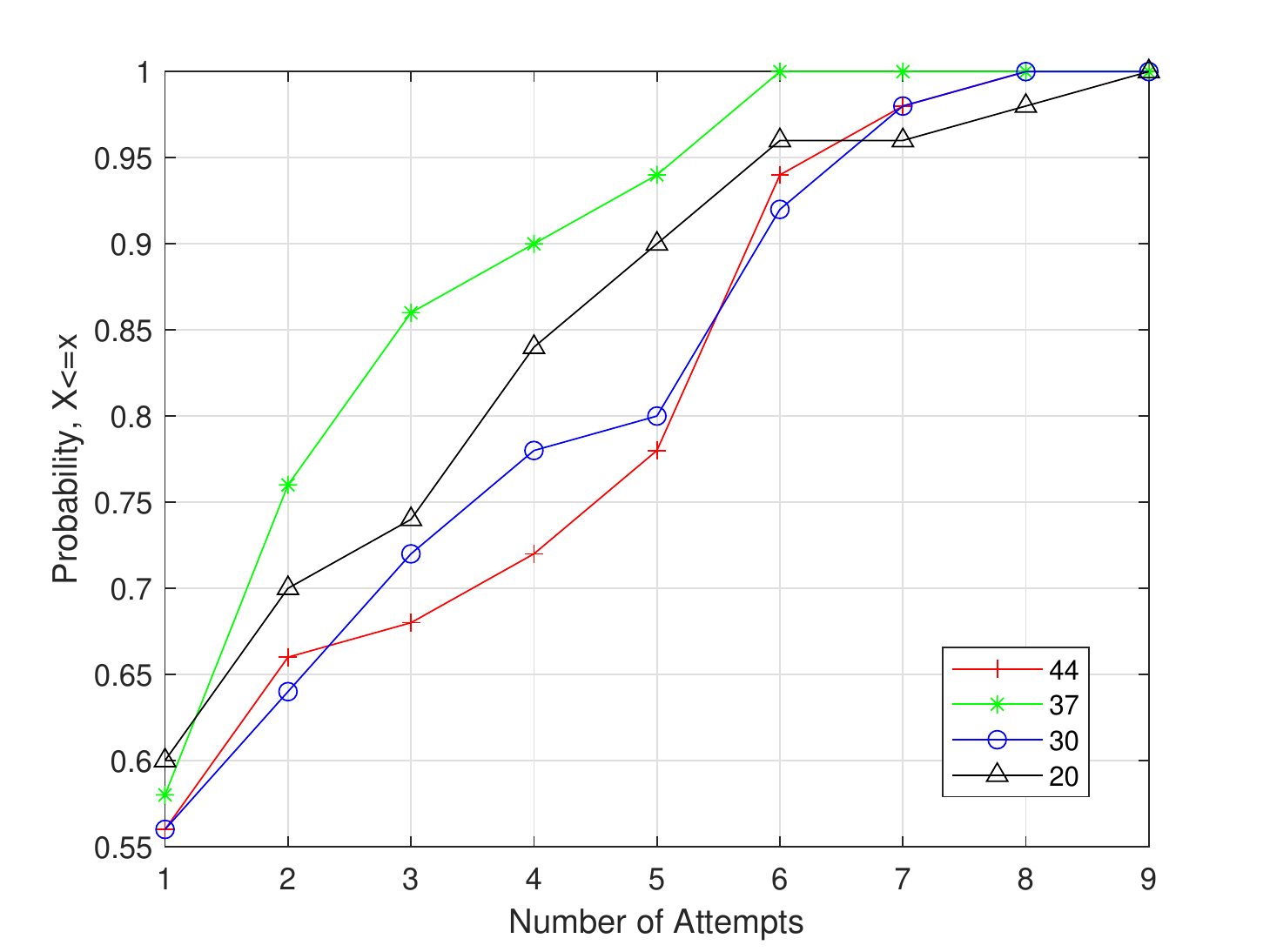}
	}
	\caption{SIA strength vs. training data size (ANN).}\vspace{1mm}
	\label{siastrength}
\end{figure}

Furthermore, we study the optimization of the inference accuracy through training
data selection in Figure~\ref{siastrength}. Selecting the right training data is
usually vital to ensure a good performance of ANN due to the
well-known over-training issue. Because the inference in case III has
been very accurate, here we only focus on the optimization of cases
I and II. For each case, we pick $k$ nearest links to the target
link, and use their signatures ($5*k$ in total) to train the ANN.
Such a treatment is based on the rationale that a closer link to the
target should possess higher correlation, and thus can provide
better training effects. So now the training data selection is
converted to deciding the optimal size of the training dataset
(i.e., the $k$). In our experiment we vary the value of $k$ (ranging
from 20 to 44) and evaluate the strength of the resulting attacks in
terms of the CDF of the number of guesses needed to find a point on
the link signature.

Figure~\ref{siastrength} plots the CDFs under various training sizes. It shows that
the security strength of SIA in both cases are sensitive to the size
of the training data. For example, the 75-percentiles in case II may
range from 2 to 5 under various training data sizes, corresponding
to a factor of $(5/2)^{50}$ difference in size of the key search
space! This observation suggests the necessity of optimizing the
training dataset in order to improve the inference accuracy, and
hence enforce the attack strength, of the SIAs. Figure~\ref{siastrength} also suggests
that the inference accuracy of the ANN is a non-monotonic function
of the training data size, and there seems to be an optimal training
data size in each case that maximizes the inference accuracy. For
example, the optimal training data sizes are 30 and 37 for case I
and case II, respectively.

\begin{figure}[H]
	\centering
	\includegraphics[scale=0.5]{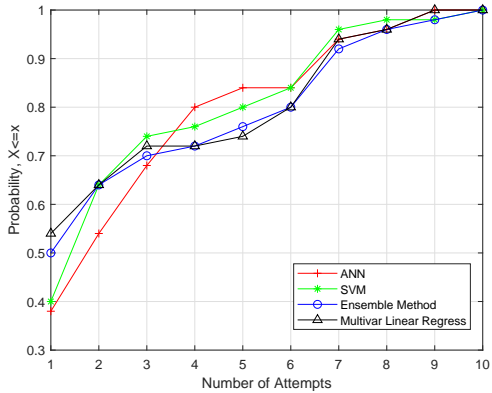}
	\caption{Comparison between different ML inference algorithms in CDF representation.}\vspace{1mm}\label{fig_ml}
\end{figure}

To test the inference performances of different ML algorithms, we utilize more ML algorithms, such as ensemble methods, support vector machine (SVM), and multivariate linear regression to launch SIAs in case I. Figure~\ref{fig_ml} plots the CDFs under different ML inference algorithms. The CDFs in this figure are also calculated based on the same 26 target links as shown in Figure~\ref{guess}, and the training data size is 44. This figures shows that more than $50\%$ points can be found in just one round by applying multivar linear regress method. In comparison, less than $40\%$ points can be found in one round by applying ANN. However, all these ML algorithms can successfully guess the truth value of each point within 10 attempts. Table II shows that statistically SVM has the highest inference accuracy, since in average, it just need 2.9 attempts to reach truth value of each point. However, when training data size becomes greater, it will not be efficient enough to launch SIA by using SVM. In this case, SVM will spend costly computation and memory resources. In addition, the adversary has to spend a lot of time to select optimal kernel and adjust the parameters in SVM.

\begin{table}[H]
	\centering
	\caption{Summary of statistics for accuracies. Note that the listed number of attempts and SD represents for each point. }
	\begin{tabular}{| c | c | c | }
		\hline
		\centering
		& Mean & Standard Deviation (SD) \\ \hline
		ANN                     & 3.04    & 2.3997         \\ \hline
		SVM                     & 2.90    & 2.4021         \\ \hline
		Ensemble Method         & 3.02    & 2.7018        \\ \hline
		Mutivar Linear Regress  & 2.94    & 2.6564        \\ \hline
	\end{tabular}
\end{table}

In our experiment, we apply several general ML algorithms to launch SIAs. How to improve the inference algorithms, and analytically decide the optimal
training data size, so that the adversary can construct the best site
survey strategy to maximize its attack strength, remains questions
to be explored in our future study.

\section{Countermeasure}\label{sec:countermeasures}
In this section, we develop a novel LSB key extraction scheme to defend against the statistical inference attacks.

\subsection{Forward-backward Cooperative Key Extraction Protocol with Helpers (FBCH)}
In conventional LSB key extraction scheme, only the link between legitimate transmitter and receiver is measured to obtain CIRs as the random series. Based on the nature of channel correlation, the adversary can effectively utilize the channel information of surrounding links to infer CIR of target link. Our experiments in Section VI demonstrate the search space of the secret key has been significantly shrunk and the inference attacks are feasible. To overcome the weakness of existing scheme, we propose a novel LSB key extraction protocol, called forward-backward cooperative key extraction protocol with helpers (FBCH). In FBCH, helpers participate in key extraction process to construct several channels, and pass the CIR information between transmitter and receiver by manipulating their transmission power. Algorithm 1 describes the procedures of FBCH.

\begin{algorithm}
    \caption{FBCH Protocol}
    \label{alg:A}
    \begin{algorithmic}
        \STATE Step 1: The Tx randomly picks $N$ helpers $H_{i}$ from all of the $M$ available relays in its transmission range.
        \STATE Step 2: The Tx broadcasts training symbols under the standard transmission power $P_{T}$, each helper $H_{i}$ receives signal from the Tx and measures the channel Tx $\to H_{i}$ to obtain the CIR $h_{tx,i}$.
        \STATE Step 3: Each helper $H_{i}$ broadcasts a training symbol under the standard transmission power $P_{H}$, the Tx and Rx receive signals from $H_{i}$ and measure channels $H_{i}\to$ Tx and $H_{i} \to$ Rx to obtain the CIRs $h_{tx, i}$, and $h_{i,rx}$, respectively.
        \STATE Step 4: The Rx broadcasts a training symbol under the standard transmission power $P_{R}$, each helper $H_{i}$ receives signal from the Rx and measures the channel Rx $\to H_{i}$ to obtain the CIR $h_{i,rx}$.
        \STATE Step 5: Each helper $H_{i}$ manipulates its transmission power to $P_{H}'$ and transmits training symbol to the Tx under the transmission power $P_{H}'$, where $P'_{H}=\frac{P_{H}*h_{i,rx}}{h_{tx, i}}$; the Tx measures the channel $H_{i}\to$ Tx to obtain the CIR $h'_{tx, i}$.
        \STATE Step 6: Each helper $H_{i}$ manipulates its transmission power again to $P_{H}''$ and transmits training symbol to the Rx under the transmission power $P_{H}''$, where $P''_{H}=\frac{P_{H}*h_{tx,i}}{h_{i, rx}}$; the Rx measures the channel $H_{i}\to$ Rx to obtain the CIR $h'_{i, rx}$.
        \STATE Step 7: The Tx and Rx utilize the summation $\sum (h_{tx,i}+h_{i,rx})$ as random series to extract secret keys.
    \end{algorithmic}
\end{algorithm}

We now detail our key extraction protocol FBCH, which consists of the following steps:

Let a transmitter Tx and a receiver Rx be the two parties that wish to extract a key; when Tx wants to establish a secret key with Rx, the Tx first
determines an integer N as the number of helpers, and randomly picks N helpers $H_{i}$ from all of the available relays in its transmission range. The number of available relays is $M$. Then the Tx broadcasts training symbols $\mathbf x_{T}$ under the standard transmission power $P_{T}$, the received signal at each helper $H_{i}$ from the Tx is given by

\begin{equation}
\mathbf y_{i,tx} = P_{T}h_{tx,i}\mathbf x_{T} + \mathbf N
\end{equation}
where $\mathbf N$ is the additive Gaussian white noise. Thus, each $H_{i}$ can measure the channel Tx $\to H_{i}$ and obtain the CIR $h_{tx,i}$.

Then each helper $H_{i}$ broadcasts a training symbol $\mathbf x_{i}$ under the standard transmission power $P_{H}$, the Tx and Rx receive signals from $H_{i}$, the received signals at the Tx and Rx are given by

\begin{equation}
\mathbf y_{tx,i} = P_{H}h_{tx,i}\mathbf x_{i} + \mathbf N
\end{equation}
and
\begin{equation}
\mathbf y_{rx,i} = P_{H}h_{i,rx}\mathbf x_{i} + \mathbf N
\end{equation}
, respectively. The Tx and Rx measure channels $H_{i}\to$ Tx and $H_{i} \to$ Rx to obtain the CIRs $h_{tx, i}$, and $h_{i,rx}$, respectively.

In the next step, the Rx broadcasts a training symbol $\mathbf x_{R}$ under the standard transmission power $P_{R}$, each helper $H_{i}$ receives signal from the Rx, the received signals at each helper $H_{i}$ is given by
\begin{equation}
\mathbf y_{i,rx} = P_{R}h_{i,rx}\mathbf x_{R} + \mathbf N
\end{equation}
Therefore, each helper $H_{i}$ can measure the channel Rx $\to H_{i}$ and obtain the CIR $h_{i,rx}$.

By using power control technology, each helper $H_{i}$ manipulates its transmission power to $P'_{H}$ and transmits the same training symbols $\mathbf x_{i}$ to the Tx again, where $P'_{H}=\frac{P_{H}h_{i,rx}}{h_{tx, i}}$; the received signal $\mathbf y_{tx,i}'$ at the Tx is given by

\begin{equation}
\begin{split}
\mathbf y_{tx,i}' &= P'_{H}h_{tx,i}\mathbf x_{i} + \mathbf N \\
&= \frac{P_{H}h_{i,rx}}{h_{tx, i}}h_{tx,i}\mathbf x_{i} + \mathbf N \\
&= P_{H}h_{i,rx} \mathbf x_{i} + \mathbf N
\end{split}
\end{equation}

When the Tx measures the channel $H_{i}\to$ Tx again, the CIR $h'_{tx, i}$ that the Tx obtains is given by

\begin{equation}
h'_{tx, i} = h_{i,rx}
\end{equation}

Again, each helper $H_{i}$ manipulates its transmission power to $P''_{H}$ and transmits the same training symbols $\mathbf x_{i}$ to the Rx, where $P''_{H}=\frac{P_{H}h_{tx,i}}{h_{i, rx}}$; the received signal $\mathbf y_{rx,i}'$ at the Rx is given by

\begin{equation}
\begin{split}
\mathbf y_{rx,i}' &= P''_{H}h_{i,rx}\mathbf x_{i} + \mathbf N \\
&= \frac{P_{H}h_{tx,i}}{h_{i,rx}}h_{i,rx}\mathbf x_{i} + \mathbf N \\
&= P_{H}h_{tx,i}\mathbf x_{i} + \mathbf N
\end{split}
\end{equation}

When the Rx measures the channel $H_{i}\to$ Rx again, the CIR $h'_{i, rx}$ that the Rx obtains is given by

\begin{equation}
h'_{i, rx} = h_{tx,i}
\end{equation}

Since the Tx and Rx obtain the CIRs $h_{tx,i}$ and $h_{i,rx}$ in step 3, respectively, they have the agreement that

\begin{equation}
\begin{split}
h_{tx,i} + h'_{tx,i} &= h_{i, rx} + h'_{i, rx} \\
&= h_{tx,i}+h_{i,rx}
\end{split}
\end{equation}

Therefore, the Tx and the Rx  are able to use the summation $\sum\limits_{i=1}^{N} (h_{tx,i}+h_{i,rx})$ as random series to extract secret key bits, where $N$ is the number of helpers. We should note that the power $P'_{H}$ and $P''_{H}$ are private and secret information of each helper $H_{i}$, which cannot be revealed to the adversary.

\subsection{Security Analysis}

\begin{figure*}[htbp]
	\centering
	\subfloat[Case I]{\includegraphics[scale=0.39]{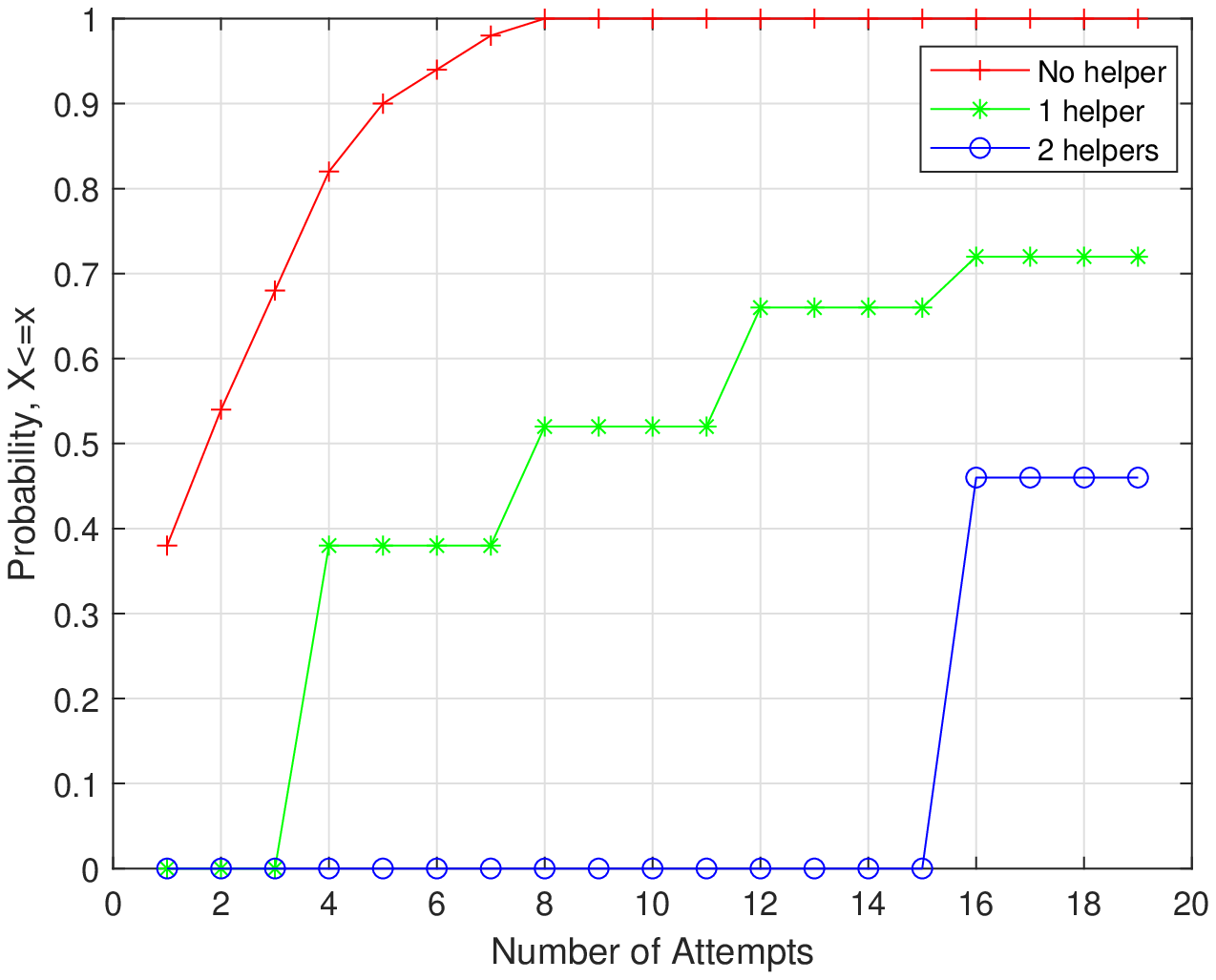}
		}
	\hfil
	\subfloat[Case II]{\includegraphics[scale=0.39]{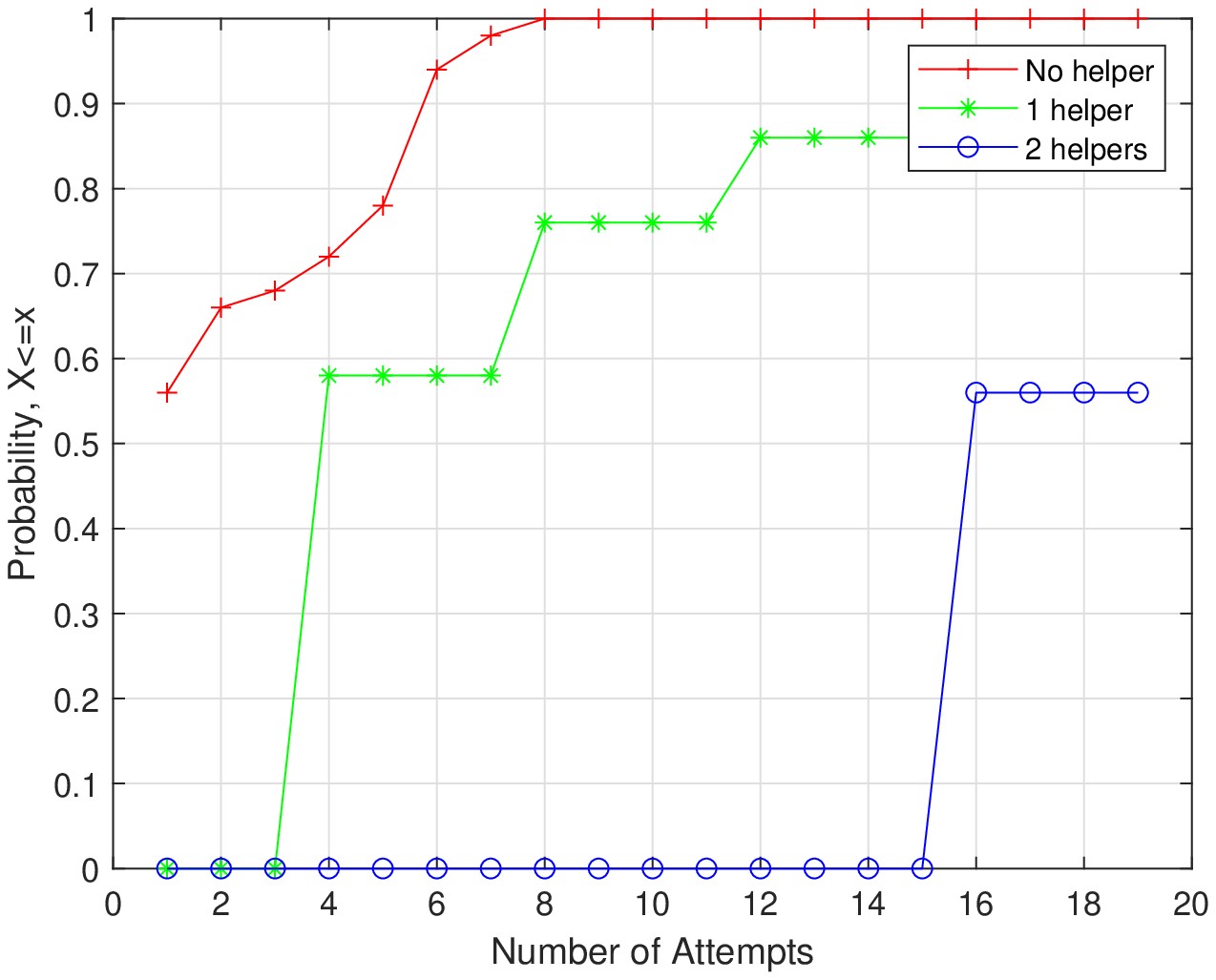}
		}
	\hfil
	\subfloat[Case III]{\includegraphics[scale=0.39]{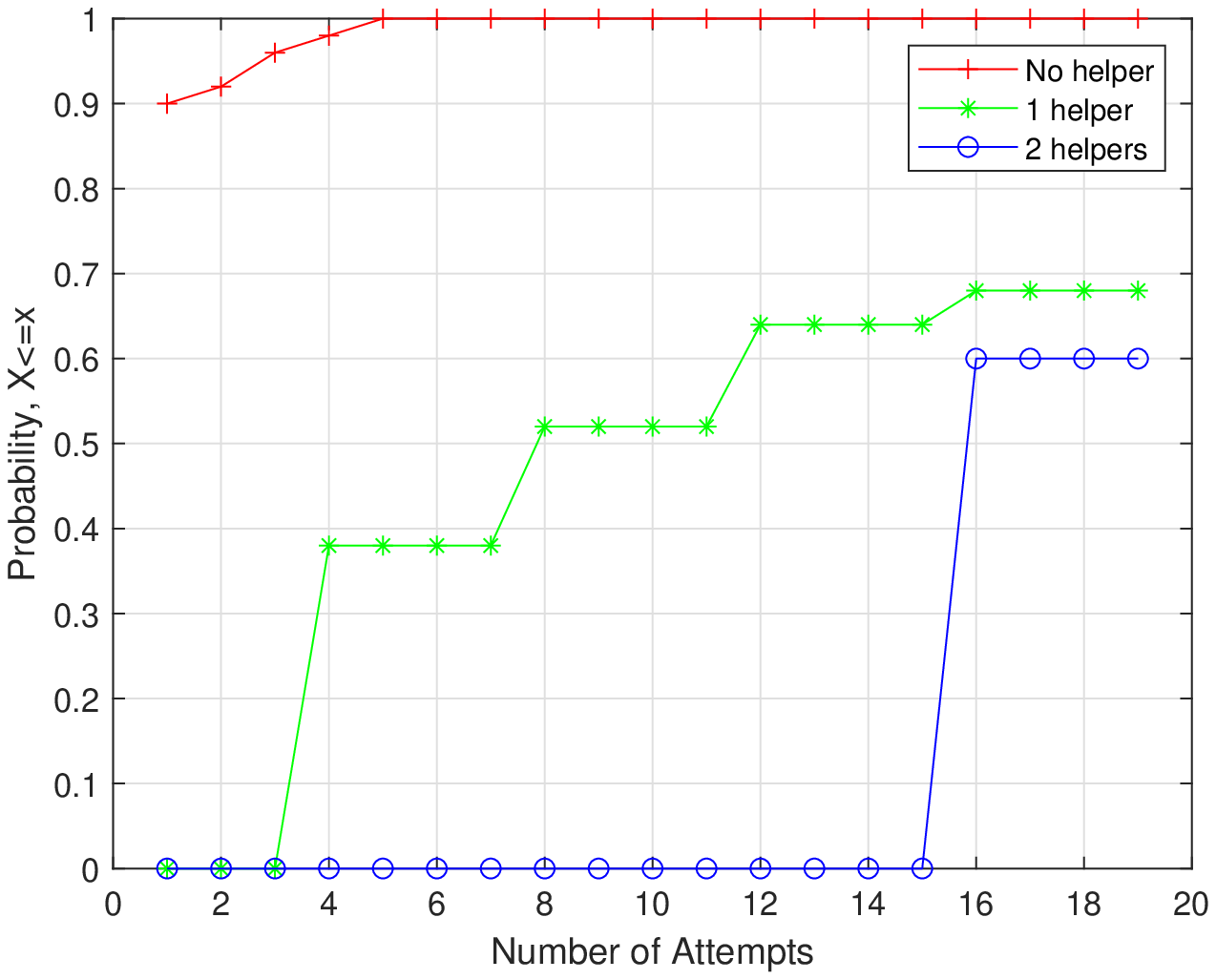}
		}
	\caption{SIA strength vs. number of helpers (SVM).}
	\label{countmeasure}
\end{figure*}

To evaluate the security of the proposed secret key extraction protocol FBCH, in this subsection, we analyze the performance of FBCH under attack models of SIAs.

First of all, in the FBCH protocol, the Tx and Rx use the summation $\sum\limits_{i=1}^{N} (h_{tx,i}+h_{i,rx})$ to extract key bits ($N$ is the number of helpers), the number $N$ is secret to the passive adversary, it is hard for the passive adversary to get this summation of CIRs for key extraction. Moreover, by introducing helper nodes, the FBCH protocol deprives the relevance between one link signature and the corresponding two locations of Tx and Rx. As a result, in the operational phase of SIAs, the ML models are not able to provide the proper group link signatures, since the adversary has no knowledge about the construction of new channels during its site survey phase. Likewise, it is hard for the adversary to infer the key bits in case II and case III of SIAs by applying ML methods. 

Second, we analyze the security strength of FBCH from the spatial randomness perspective. By launching SIAs, the adversary has the ability to infer the link signature of any channel between any two locations. If there exists at least one attack node that is in close proximity to each helper node (here the "close" means that the distance between attack node and helper is smaller than $\lambda /2$), the strong correlation will lead to the accurate inference of link signature on each legitimate link. Therefore, the close distance between attack nodes and helpers will threaten the security of FBCH protocol. We will study the probability that an attack node is placed very close to a legitimate node by lucky coincidence.

Since the distributions of attack nodes and legitimate nodes (i.e., Tx, Rx and helpers) are fully random, the number of attack nodes that are close to each legitimate node follows the Poisson distribution. Therefore, we can apply the spatial Poisson point process to analyze this probability as follows:

The Poisson point process is defined in the plane $\mathbf{R^{2}}$. And we consider a circular area $B_{i} \subset \mathbf{R^{2}}$ ($i=1...N$, and $N$ is the number of legitimate nodes) for one legitimate node, which takes the legitimate node as the center and $r$ as the radius. We treat the attack nodes as points in the plane $\mathbf{R^{2}}$. The number of points of a point process $X$ existing in this area $B_{i}$ is a random variable, denoted by $X(B_{i})$. The points belong to a Poisson process with parameter $\lambda>0$, then the probability of $k$ points existing in $B_{i}$ is given by

\begin{equation}
P\{X(B_{i})=k\} = \frac{(\lambda|B_{i}|)^k}{k!} \cdot exp(-\lambda |B_{i}|),
\end{equation}
where $|B_{i}|$ denotes the area of $B_{i}$, and $|B_{i}|=\pi r^{2}$; $\lambda$ is the density of points in the plane $\mathbf{R^{2}}$.

Given by Eq.(30), the probability that at least one point existing in $B_{i}$ is given by 

\begin{equation}
\begin{split}
1-P\{X(B_{i})=0\} &=1-exp(-\lambda |B_{i}|) \\
&=1-exp(-\lambda \pi r^{2})
\end{split}
\end{equation}

To launch SIAs successfully, there should be at least one attack node existing in $B_{i}$ for every legitimate node. Then the total probability $P$ that at least one attack node existing in $B_{i}$ for each legitimate node is given by

\begin{equation}
\begin{split}
P &=\prod_{i=1}^{N}(1-exp(-\lambda |B_{i}|)) \\
&=(1-exp(-\lambda \pi r^{2}))^{N},
\end{split}
\end{equation}
where $N$ is the number of legitimate nodes.

It can be observed from Eq.(32) that when the number of helpers $N$ becomes large or the density of attack nodes $\lambda$ becomes small, the probability $P$ becomes small.

In particular, if $N$ gets very large, we obtain 

\begin{equation}
P =\lim_{N \to \infty}(1-exp(-\lambda \pi r^{2}))^{N} = 0
\end{equation}

Given by Eq.(33), we can observe that when we pick large enough number of helpers (i.e., $N \to \infty$), the probability that there exists at least one attack node in $B_{i}$ for each legitimate node is 0, which implies that there is no possible for the adversary to launch SIAs successfully.

Obviously, FBCH will exponentially increase the overhead of communication in the number of helpers during the key extraction process. Nevertheless, it is an effective countermeasure solution to defend against the SIAs.

\subsection{Numerical Results}

To illustrate the security strength of FBCH, in this subsection we study the performance of statistical inference attacks (SIAs), which we proposed in previous sections, to the new key extraction scheme. Our experiments are based on the same setup as in Section VI-B: we use the same CRAWDAD dataset and launch SIAs in 3 cases. As pointed out in Section VI, support vector machine (SVM) has the highest inference accuracy in the previous experiments. To make a fair comparison, we only use SVM as the ML inference method to launch SIAs.

In this experiment, we first pick $N$ nodes from CRAWDAD as the legitimate Tx, Rx and helpers, respectively. Furthermore, we randomly pick several links as the transmitter-helpers (Tx-H) links and helpers-receiver (H-Rx) links. Then we use the summations of CIRs of these several links, $h+h'+...$, as ground truth to extract secret keys. Since in each case of SIAs, the adversary has no knowledge about the helper’ selection, he has to infer CIRs on all links according to each potential helper (the location of each helper is known to the adversary). For case I SIA, the adversary uses all remaining links in CRAWDAD to infer the CIRs of potential Tx-H and H-Rx links and calculate the summation of these CIRs. Then he uses this summation to guess secret keys, as we mentioned in Section III-B. Likewise, for case II SIA, the adversary attempts to infer each potential Tx-H and H-Rx links using links that share the same receiver but have a different transmitter. And for case III SIA, the adversary randomly picks helpers and infer potential Tx-H and H-Rx links (the locations of Tx, Rx, and each helper are known).

To obtain the statistical view about the security strength of the proposed protocol FBCH, we study the percentage of bits in a secret key string that can be inferred, as a function of the number of attempts. Figure~\ref{countmeasure} compares the CDF (cumulative distribution function) of the number of guesses needed per point under 3 SIA cases. In each attack case, we vary the number of helpers and use different size of training data to measure the security strength of the proposed scheme. This figure shows that the adversary needs more attempts to find true key, and FBCH can exponentially amplify the the adversary's search space. For example, in case I SIA, the adversary can find $94\%$ points in just 6 rounds when Tx and Rx use conventional LSB key extraction scheme. On the contrary, $38\%$ points are found in 6 rounds when there is 1 helper, while $0$ point is found in 6 rounds when there are 2 helpers. In particular, FBCH can significantly increase the adversary's search space and effectively prevent the SIA in Case III. For example, when Tx and Rx use conventional LSB key extraction scheme, the adversary can find $90\%$ true bits in secret key in just one round, while they cannot find any true bit in 14 rounds when two or more helpers involve.

\section{Conclusion}\label{sec:conclusion}
In summary, the formal theoretical analysis in channel correlations have been done  relying on both outdoor and indoor models. Following the machine learning (ML) framework, we have studied empirical
statistical inference attacks against LSB key extraction. Different
from prior analytical work that assumes a link-correlation model,
our study roots from empirically measured channel data and does not
rely on any assumption on the link correlation. We applied several
ML-based methods to launch SIA against LSB key extraction under
various scenarios, and evaluated the effectiveness of these attacks
based on the CRAWDAD dataset. Our finding has verified the
existence of correlation between neighboring links in realistic
environments, and also showed that such correlation can be practically
exploited by ML algorithms to undermine
the security strength of PHY-layer security measures.
Upon investigation, we proposed a countermeasure against the statistical inference attacks called forward-backward cooperative key extraction protocol with helpers (FBCH). Our experiments verify that FBCH is more robust under the statistical inference attacks.

\section*{Acknowledgment}
This research work is partially supported by the National Science Foundation under Grants CNS-1837034, CNS-1745254, CNS-1659965, CNS-1659962, CNS-1460897 and DGE-1623713. Any opinions, findings, and conclusions or recommendations expressed in this material are those of the authors and do not necessarily reflect the views of the National Science Foundation.




\bibliographystyle{IEEEtran}
\bibliography{citations}

\begin{IEEEbiography}[{\includegraphics[width=1in,height=1.2in,clip,keepaspectratio]{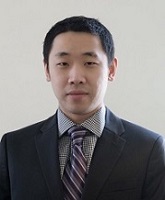}}]
{Rui Zhu} is a Ph.D. candidate in the Dept. of Computer Science and Technology at Oakland University, Rochester, MI, USA. He received the B.S. and M.S. degree from Beijing Jiaotong University, China, and Valparaiso University, USA, respectively. He worked as a system development manager in rTrail LLC. and was a Software Engineer intern at Apple Inc. His current research interests include PHY-layer security in wireless network and millimeter-wave communication security.
\end{IEEEbiography}


\begin{IEEEbiography}[{\includegraphics[width=1in,height=1.25in,clip,keepaspectratio]{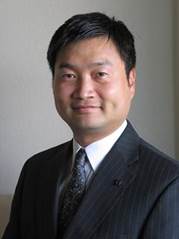}}]
{Tao Shu} received the B.S. and M.S. degrees in electronic engineering from the South China University of Technology, Guangzhou, China, in 1996 and 1999, respectively, the Ph.D. degree in communication and information systems from Tsinghua University, Beijing, China, in 2003, and the Ph.D. degree in electrical and computer engineering from The University of Arizona, Tucson, AZ, USA, in 2010. He is currently an Assistant Professor in the Department of Computer Science and Software Engineering at Auburn University, Auburn, AL. His research aims at addressing the security, privacy, and performance issues in wireless networking systems, with strong emphasis on system architecture, protocol design, and performance modeling and optimization.
\end{IEEEbiography}

\begin{IEEEbiography}[{\includegraphics[width=1in,height=1.25in,clip,keepaspectratio]{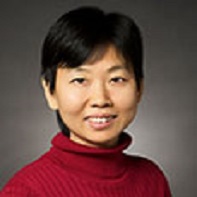}}]
{Huirong Fu} (M’01) received the Ph.D. degree from Nanyang Technological University, Singapore, in 2000. She is currently a Professor with the Department of Computer Science and Engineering, Oakland University, Rochester, MI, USA, where she joined as an Assistant Professor in 2005. Previously, she was an Assistant Professor with North Dakota State University, Fargo, ND, USA, for three years, and she was a Postdoctoral Research Associate with Rice University, Houston, TX, USA, for two years. As a Lead Professor and the Principal Investigator for several projects funded by the National Science Foundation, she has been actively conducting research in the areas of networks, security, and privacy.
\end{IEEEbiography}




\end{document}